\documentclass[useAMS,usenatbib]{mn2e}
\usepackage{epsfig}

\newcommand{\beq}{\begin{equation}}
\newcommand{\eeq}{\end{equation}}

\def\lsim{\mathrel{\lower2.5pt\vbox{\lineskip=0pt\baselineskip=0pt
           \hbox{$<$}\hbox{$\sim$}}}}
\def\gsim{\mathrel{\lower2.5pt\vbox{\lineskip=0pt\baselineskip=0pt
           \hbox{$>$}\hbox{$\sim$}}}}
\def\Lsun{\hbox{L$_{\odot}$}}
\def\Msun{\hbox{M$_{\odot}$}}

\input epsf

\title[Radio-mm-FIR photometric redshifts of SHADES sources] 
{The SCUBA HAlf Degree Extragalactic 
Survey (SHADES) -- IV: Radio-mm-FIR photometric redshifts}
\author[Aretxaga et al.]
{Itziar Aretxaga$^1$, David H. Hughes$^1$, Kristen Coppin$^{2,3}$, 
Angela M.J. Mortier$^{4,5}$,
\newauthor
 Jeff Wagg$^{6,1}$,   James S. Dunlop$^4$, Edward L.  Chapin$^3$,   
Stephen A. Eales$^7$, Enrique 
\newauthor 
Gazta\~naga$^{8,1}$,  Mark Halpern$^3$,
Rob J. Ivison$^9$, Eelco van Kampen$^{10}$, Douglas Scott$^3$, 
\newauthor
Stephen Serjeant$^{11}$, Ian Smail$^2$, Thomas Babbedge$^{12}$, 
Andrew J. Benson$^{13}$, 
  \newauthor
Scott Chapman$^{13}$, 
David
L. Clements$^{12}$, Loretta Dunne$^{14}$, Simon Dye$^{7}$, 
Duncan Farrah$^{15}$, 
\newauthor
Matt J. Jarvis$^{16,17}$, Robert G. Mann$^{4}$, 
Alexandra Pope$^3$, 
Robert Priddey$^{18}$, 
\newauthor
Steve Rawlings$^{17}$, Marc Seigar$^{19}$, Laura Silva$^{20}$, 
Chris Simpson$^{21}$, and Mattia Vaccari$^{22,12}$.
\\
$^1$Instituto Nacional de Astrof\'{\i}sica, \'Optica y Electr\'onica
(INAOE), Aptdo. Postal 51 y 216, 72000 Puebla, Pue., Mexico  \\
$^2$  Institute for Computational Cosmology, Durham University, South Road, Durham DH1 3LE, UK\\
$^3$ Department of Physics \& Astronomy, University of British
Columbia,
6224 Agricultural Road, Vancouver, B.C., V6T 1Z1, Canada\\
$^4$ SUPA (Scottish Universities Physics Alliance), 
Institute for Astronomy, University of Edinburgh, Blackford Hill,
Edinburgh, EH9 3HJ, UK\\
$^5$ Center for Astrophysics and Planetary Science, School of
Physical Sciences, University of Kent, Canterbury CT2 7NR, UK\\
$^6$  National Radio Astronomy Observatory, P.O. Box O, Socorro, NM 87801, 
USA\\
$^{7}$ Cardiff School of Physics and Astronomy, Cardiff University 5, 
The Parade, Cardiff, CF24 3YB, UK\\ 
$^8$ Institut d'Estudis Espacials de Catalunya, IEEC/CSIC, c/ Gran
Capit\`a
2-4, 08034 Barcelona, Spain\\
$^9$ UK ATC, Royal Observatory, Blackford Hill, Edinburgh, EH9 3HJ, UK\\
$^{10}$ Institute for Astrophysics, University of Innsbruck, 
Technikerstr. 25, A-6020, Innsbrusk, Austria.\\
$^{11}$ Astrophysics Group, Department of Physics, The Open University,
Milton Keynes, MK7 6AA, UK\\
$^{12}$ Astrophysics Group, Blackett Laboratory, Imperial College, Prince Consort Rd., London SW7 2BW, UK\\ 
$^{13}$ Caltech, 1200 E. California Blvd., Pasadena, CA 91125-0001, USA\\
$^{14}$ The School of Physics and Astronomy, University of Nottingham, 
University Park, Nottingham, NG7 2RD, UK\\
$^{15}$ Department of Astronomy, Cornell University, Space Sciences Building, Ithaca, NY 14853, USA\\
$^{16}$ Centre for Astrophysics Research, Science \& Technology Research  
Institute, University of Hertfordshire, Hatfield, AL10 9AB, UK\\
$^{17}$ Department of Astrophysics, Denys Wilkinson Building, Keble Road, Oxford, OX1 3RH, UK\\
$^{18}$ Department of Physics, Astronomy \& Mathematics, University of Herdfordshire, College Lane, Hatfield, Hertfordshire AL10 9AB, UK\\
$^{19}$ Center for Cosmology, Dept. of Physics \& Astronomy, Univ. of California, Irvine, 4129 Frederick Reines Hall, Irvine, CA 92697-4575, USA\\
$^{20}$ Osservatorio Astronomico di Trieste, Via Tiepolo 11, I-341311, Trieste, 
Italy\\
$^{21}$ Astrophysics Research Institute,
Liverpool John Moores University, Twelve Quays House, Egerton Wharf,
Birkenhead CH41 1LD, UK\\
$^{22}$  
Department of Astronomy, University of Padova, Vicolo dell'Osservatorio
2, I-35122, Padova, Italy\\
}

\begin{document}

\date{}

\pagerange{} \pubyear{}

\maketitle

\label{firstpage}

\begin{abstract}
  We present the redshift distribution of the SHADES galaxy population
  based on the rest-frame radio-mm-FIR colours of 120 robustly
  detected 850$\mu$m sources in the Lockman Hole East (LH) and Subaru XMM-Newton
  Deep Field (SXDF). The redshift distribution derived
  from the full SED information is shown to be narrower than that
  determined from the radio--submm spectral index, as more photometric
  bands contribute to a higher redshift accuracy.  The
  redshift distribution of sources derived from at least two photometric bands
  peaks at $z\approx 2.4$ and has a near-Gaussian distribution, with
  50~per cent (interquartile range) of sources at $z=1.8-3.1$.  
  We find a statistically-significant difference between the measured
redshift distributions in the two fields; the SXDF peaking at a
slightly lower redshift (median $z\approx 2.2$) than the LH (median
$z\approx 2.7$), which we attribute to the noise-properties of the
radio observations. We demonstrate however that there could also
be field-to-field variations that are consistent with the measured
differences in the redshift distributions, and hence, that the incomplete 
area observed by SHADES with SCUBA, despite being the largest sub-mm survey
to date, may still be too small to fully characterize the bright
sub-mm galaxy population.
Finally we present 
a brief comparison with the predicted, or assumed, redshift
distributions of sub-mm galaxy formation and evolution models, and 
we derive the contribution of these SHADES sources and the general 
sub-mm galaxy population
to the star formation-rate density at different epochs.

\end{abstract}


\section{Introduction}

The SCUBA HAlf Degree Survey (SHADES, Dunlop 2005, Mortier et
al. 2005) was originally designed with the aim of characterizing the
star-formation history (Hughes et al. 2002) and clustering properties
(van Kampen et al. 2005) of the bright-end of the luminous
dust-enshrouded galaxy population.  To achieve these goals we mapped two
regions of the sky centered on the Lockman Hole East (LH) and Subaru
XMM-Newton Deep Field (SXDF) with the Submillimetre Common-User
Bolometer Array (SCUBA, Holland et al. 1999). With a proposed
1$\sigma$ sensitivity of 2~mJy at 850$\mu$m the complete survey was
predicted to identify a statistically robust sample of $\sim 200$
galaxies, with sufficient radio to FIR ancillary data to help identify
optical/IR counterparts and derive spectroscopic/photometric
redshifts. This redshift information is essential for determining 
the star formation and
clustering properties for the whole population of ultraluminous
dust-enshrouded galaxies. SCUBA was de-commissioned in mid-2005 having
covered $\sim 40$~per cent of the originally-proposed area of the
SHADES\footnote{
The complete 1800 sq. arcmins SHADES area towards the LH and
the SXDF has recently been surveyed at the JCMT at 1.1mm with AzTEC 
(Wilson et al. 2004), a continuum camera destined for the 50-m Large 
Millimetre Telescope (Serrano et al. 2006).  These 
AzTEC data are currently 
being analysed and the results will be presented elsewhere.
}.

Paper~I of this series (Mortier et al. 2005) describes the survey
motivation, strategy and the philosophy adopted for the analysis. 
Paper~II (Coppin et al. 2006) presents the catalogue and number
counts derived from the 850$\mu$m sources. Paper~III (Ivison et
al. 2007) describes the identification of radio and mid-IR
counterparts of these sources. This paper (IV) constructs the redshift
distribution derived from the radio-mm-FIR
photometry of the SHADES sources based on a compilation of the
850$\mu$m and 450$\mu$m
SCUBA data (Coppin et al. 2006), 1.4GHz Very Large Array photometry
(Ivison et al. 2007) and other  previously published mm to FIR photometric
observations towards these fields. A study of the mid-IR
to optical properties of the SHADES population, and further constraints
on the photometric redshifts of the sources, 
will be published elsewhere (Clements et al. 2007,
Dye et al. 2007, Serjeant et al. 2007 ). A 
spectroscopic study of a sub-sample of SHADES sources with 
identified optical/IR counterparts (Blain  et al. 2007) will also 
provide an important comparison of spectroscopic and 
photometric redshifts.

The cosmological parameters adopted throughout this paper are 
$H_0=71$~km\,s$^{-1}$\,Mpc$^{-1}$, $\Omega_{\rm M}=0.27$,
$\Omega_{\Lambda}=0.73$.

\section{Photometric redshifts}
Despite having mapped only $\sim40$~per cent of the planned
0.5 sq. degree area, SHADES remains the largest extragalactic 
sub-mm survey to date. The
difficulties of following-up such large areas at other wavelengths,
and hence the inhomogeneity of the multi-wavelength data, implies that
the same photometric redshift technique cannot be applied to all
sources.  This section has been divided in two subsections: the first
(\S~2.1) deals with the consideration of 850$\mu$m and 1.4GHz
photometry which is available for all sources, and the use of the
sub-mm--radio spectral-index as a diagnostic of redshift; and the
second (\S~2.2) describes the inclusion of additional photometry at 70
to 450$\mu$m which is sufficiently sensitive to place important constraints
on the radio-mm-FIR photometric redshifts for only a few tens of
sources. In both subsections we make a brief introduction to the
techniques used, the estimated uncertainties 
found when comparing photometric and
spectroscopic redshifts for similar sub-mm galaxies, the results from
the application of the techniques to individual SHADES sources, and
the combined redshift distributions derived for the entire SHADES
population.

\subsection{1.4GHz/850$\mu$m spectral index}

\subsubsection{Techniques and accuracies}
\label{sec:compaCY}

One of the simplest redshift-indicators for the sub-mm galaxy population is
that formed by the ratio of the flux densities 
at 1.4GHz and 850$\mu$m.  These wavebands
trace the tight correlation between radio continuum emission, which is
dominated by synchrotron radiation from supernova remnants, and
thermal emission from warm dust heated by young stars (Helou, Soifer
\& Rowan-Robinson 1985, Condon 1992, Yun, Reddy \& Condon 2001).  This
redshift indicator was systematically studied by Carilli \& Yun (1999,
2000), and has been subsequently revised for different sub-mm galaxy
sub-populations (Dunne, Clements \& Eales 2000, Rengarajan \& Takeuchi
2001). The 1.4GHz to 850$\mu$m flux-density ratio, or a spectral index derived
from it, increases monotonically with redshift, with some degeneracy
due to the variety of radio synchrotron-slopes and mm dust-emissivity
indices present in the ISM of those local galaxies used to define the
relationship. Additionally there exists a level of degeneracy between the
temperature of the dust generating the rest-frame FIR luminosity (and
hence sub-mm flux) and the redshift.  Regardless, by adopting a library 
of local galaxy templates, and accepting the intrinsic dispersion in their SEDs,
the 1.4GHz to
850$\mu$m flux-density ratio still provides a crude but useful estimation of the
redshift.  This indicator becomes relatively insensitive to redshift
beyond $z\sim 3$, as the 850$\mu$m filter starts to sample the
flattening of the spectral energy distribution (SED) towards the
rest-frame FIR peak, whilst still providing a powerful discriminant
between low-redshift ($z<2$) and high-redshift ($z>2$) objects.

\begin{figure}
\hspace*{-1.1cm}
\epsfig{file=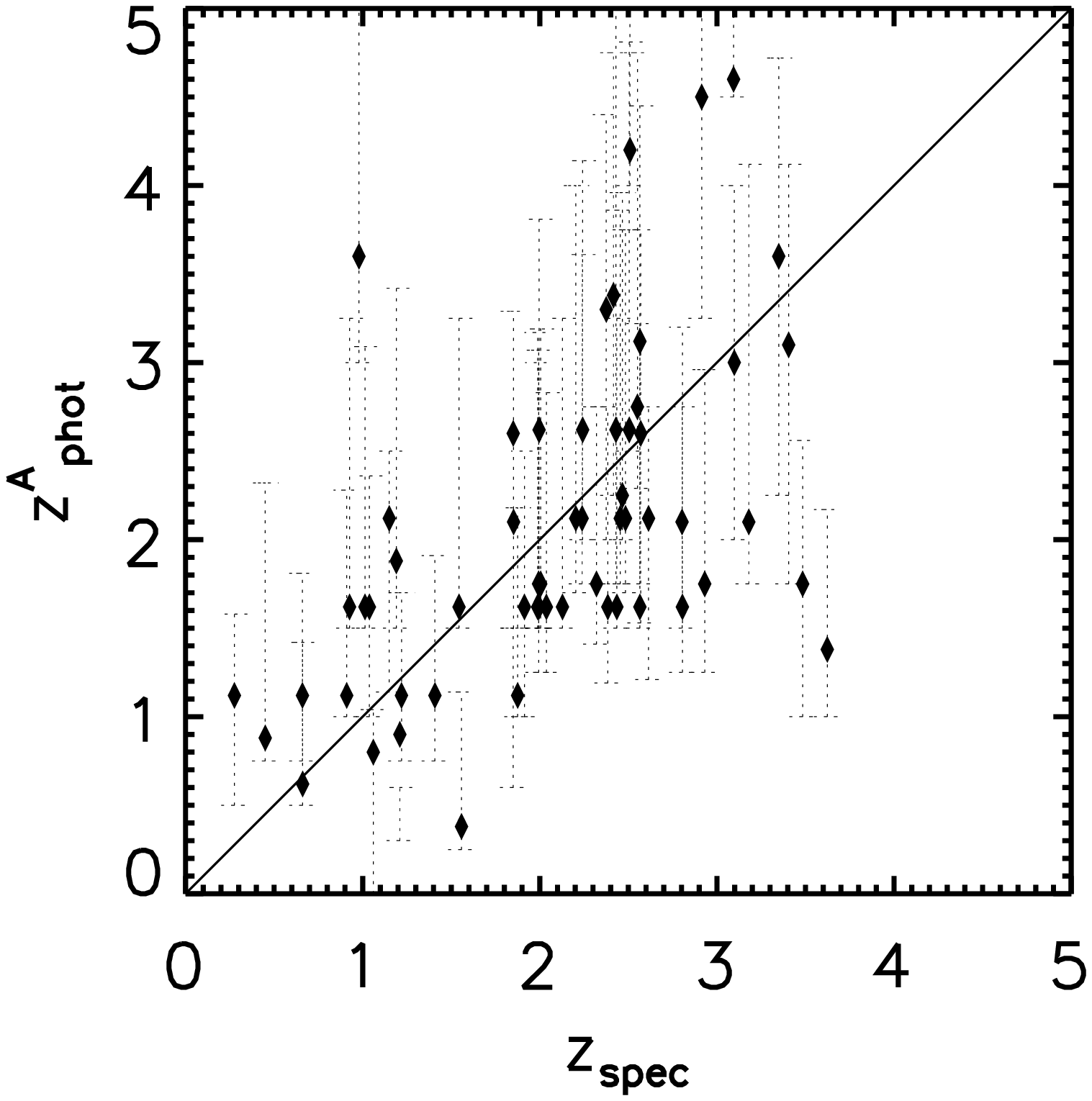,width=1.1\hsize,angle=90} 
\hspace*{-0.7cm} 
\epsfig{file=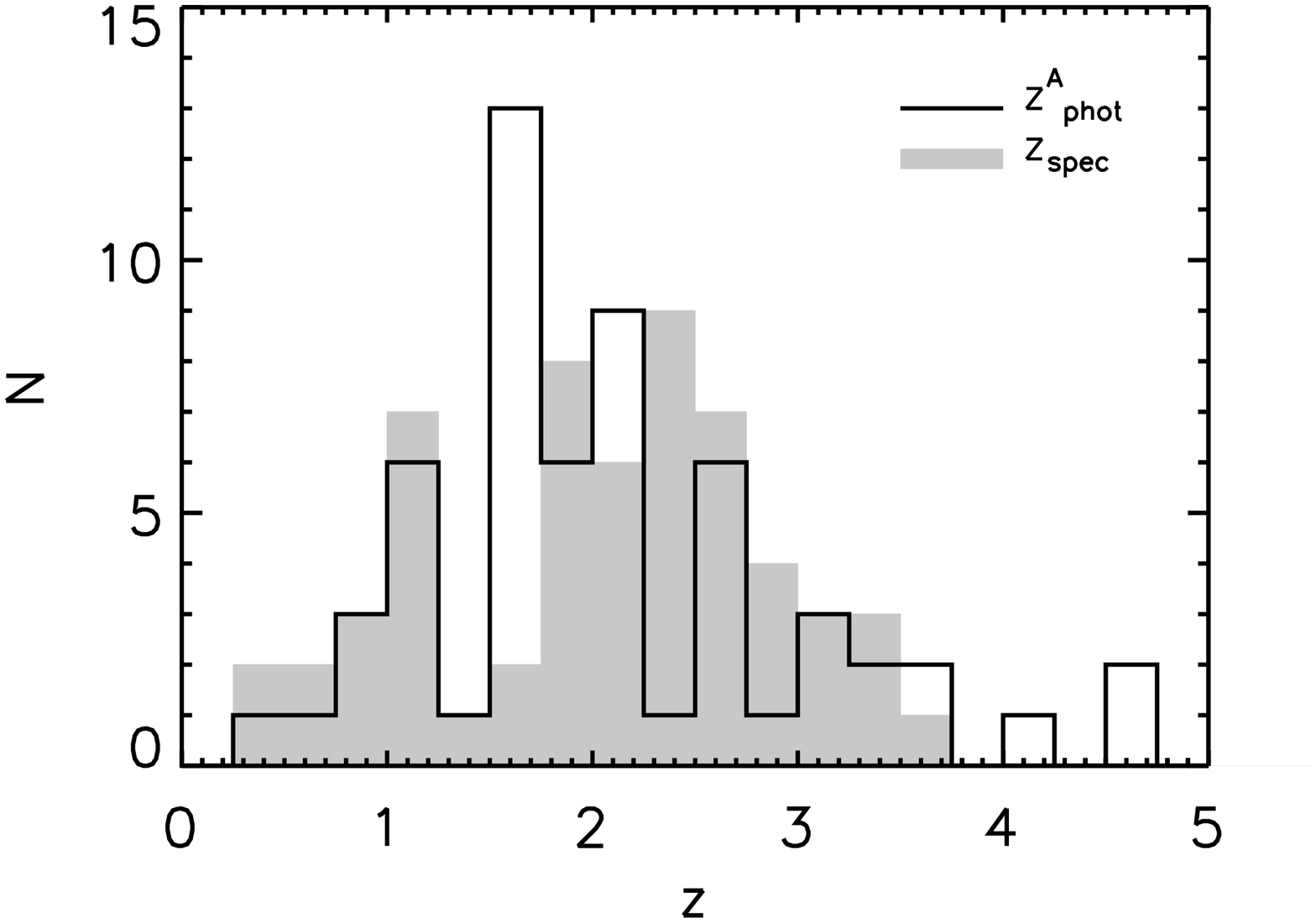,width=0.85\hsize,angle=90}
%\hspace*{-1.1cm}
%\epsfig{file=SEDall_zcum.ps,width=0.40\hsize,angle=90} 
%\hspace*{-0.7cm} 
%\epsfig{file=SEDall_zcum_smooth.ps,width=0.40\hsize,angle=90}
%\end{figure}
%\begin{figure}
  \caption{ 
    {\bf (Top)} Comparison of spectroscopic and photometric redshifts
    derived from the 1.4GHz/850$\mu$m spectral index for a sample of
    58 sub-mm galaxies with undisputed
    radio/optical/IR identifications,
    and spectroscopic redshifts derived from 2 or more lines 
   (Aretxaga et al. 2006). The
    error bars represent 68\% confidence intervals in the
    determination of the redshift. The r.m.s. scatter of the relation 
    $z_{\rm spec} - z_{\rm phot}^{\rm A}$displayed   is 0.8. {\bf (Bottom)} 
    Histogram distribution of
    the spectroscopic and photometric redshifts represented in the top
    panel, which illustrates the success in recovering the redshift 
    distribution of the sample. }
\label{fig:zzCY}
\end{figure}

We shall discuss the 1.4GHz/850$\mu$m spectral index following two
different prescriptions: (a) the single-template maximum-likelihood technique
originally designed by Carilli \& Yun (1999, 2000), denoted as $z_{\rm
phot}^{\rm CY}$; and (b) a maximum likelihood technique which simultaneously
fits the 20 local templates of starbursts, ULIRGs and AGN used by
Aretxaga et al. (2003, 2005), denoted as $z_{\rm phot}^{\rm A}$.

The success of any photometric-redshift technique is measured by the
accuracy with which it can predict the individual redshifts for a
sample of representative galaxies with known redshifts, which have not
been used to define the method. Aretxaga, Hughes \& Dunlop (2006) have
previously assessed the accuracy of the above two 1.4GHz/850$\mu$m
photometric redshift indicators. Based on this study, we show in
figure\,1 a comparison of spectroscopic and photometric-redshifts for
58 sub-mm/mm selected galaxies, complemented with a few objects
selected at optical/FIR wavelengths, which have published optical/IR
or CO spectroscopic redshifts and accompanying radio-FIR
photometry. We will refer to this dataset as the `comparison sample'
hereafter.  This comparison study shows that the 
$z^{\rm A}_{\rm  phot}$ prescription has a mean accuracy 
$\Delta z \equiv <|z^{\rm
  A}_{\rm phot}-z_{\rm spec}|> \approx 0.65$ over the whole redshift
interval, when one selects a robust sub-sample of objects with
unambiguous optical/IR/radio counterparts and spectroscopic redshifts
derived from the identification of two or more spectral lines.  For
the same robust sample of objects, $z^{\rm CY}_{\rm phot}$ has
systematically larger errors, $\Delta z \approx 0.9$.  This sample
does not include powerful radio-loud AGN, for which the template SEDs
used in the photometric redshift analysis are not appropriate. The
r.m.s. of the relation is $< ( z^{\rm A}_{\rm phot}-z_{\rm spec})^2
>^{1/2} \approx 0.8$.  Restricting the analysis only to those galaxies
with CO spectroscopic redshifts, the measured accuracy is $\Delta z
\approx 0.6$ and has an r.m.s of 0.8 at $0\le z\le4$.  
The precision degrades as the
redshift increases, as expected from the 1.4GHz/850$\mu$m spectral
index, which flattens beyond $z=3$ (Carilli \& Yun 2000), leading to a
measured $\Delta z \approx 1.0$ at $3\le z \le 4$. Using all objects
with published photometry and spectroscopic redshifts, regardless of
whether the associations that lead to the spectroscopic redshift are
unambiguous or not, the overall accuracy over the $0\le z \le 4$
regime degrades to $\Delta z \approx 0.8$ (see Aretxaga et al. 2006,
figure 1).

%In what follows, in this subsection, 
%we will use the two techniques: the original Carilli
%\& Yun (2000) method ($z^{\rm CY}_{\rm phot}$) for reference, since it
%is widely used in the literature, and the variation based on the
%collection of templates used in Aretxaga et al. (2005) as an
%alternative, ($z^{\rm A}_{\rm phot}$), that gives slightly more
%accurate results.

\subsubsection{The redshifts of SHADES sources}

The radio counterparts adopted for the photometric redshift
calculations of SHADES sources are the secure sample detected within
8~arcsecs of the sub-mm position (Ivison et al. 2007), with a
chance-association probability between the radio and sub-mm source of
$P<0.05$. We have accepted some additional counterparts when a
robustly-detected radio source is still within 10~arcsec of the sub-mm
centroid and, additionally, a 24$\mu$m counterpart is associated with
this radio identification. These extra radio counterparts are marked
in the notes provided for each sub-mm source (see
tables~\ref{tab:CY_LH} and \ref{tab:CY_SXDF}), where we have
calculated the corresponding $P$-value of the radio association, which
remains lower than 0.08.  The 34 radio sources adopted as counterparts
of SHADES galaxies in the LH field, and the 35 
radio sources in the SXDF have a
combined chance association $P\approx1.6$, and thus we expect to have
incorrectly associated $\sim 1$ of the SHADES sub-mm sources with a
projected radio source.

For both techniques the error bars of the photometric redshifts
were derived by bootstrapping on the reported photometric and
calibration errors (Coppin et al. 2006, Ivison et al. 2007), and are
defined as the 68\% confidence interval of the resulting redshift
probability distribution.  The photometric error distributions used
for the 850$\mu$m photometry were derived by de-boosting the measured
flux densities of the SCUBA sources. A de-boosting correction is 
necessary to provide a more accurate
estimate of the flux of low S/N blank-field sources in sub-mm surveys, 
where the
counts are typically very steep and faint galaxies can be statistically boosted
above the 
nominal detection threshold. These errors
are often non-Gaussian (see figure~5 in
Coppin et al 2006). The 1.4GHz flux densities do not need to be de-boosted,
since this correction is dependent on the area defined by
the search radius in identifying the source. 
In the case of finding associations within $\lsim 8$~arcsec 
radius around a
known object, this is
negligible.  The error distributions for the 1.4GHz flux densities were
assumed to be Gaussian. In the case of $z_{\rm phot}^{\rm CY}$, the
error estimated by Carilli \& Yun (2000), to allow for a difference in
templates, is added in quadrature to the errors derived by
bootstrapping the photometry.

The probability distribution calculated for each source in this
  manner typically have a single peak, which broadens as the most probable
  redshift of the source increases.  Figure~2 shows an example of a 
  typical solution derived from the use of 850$\mu$m and 1.4GHz photometry.

\begin{figure}
\hspace*{-1.1cm}
\epsfig{file=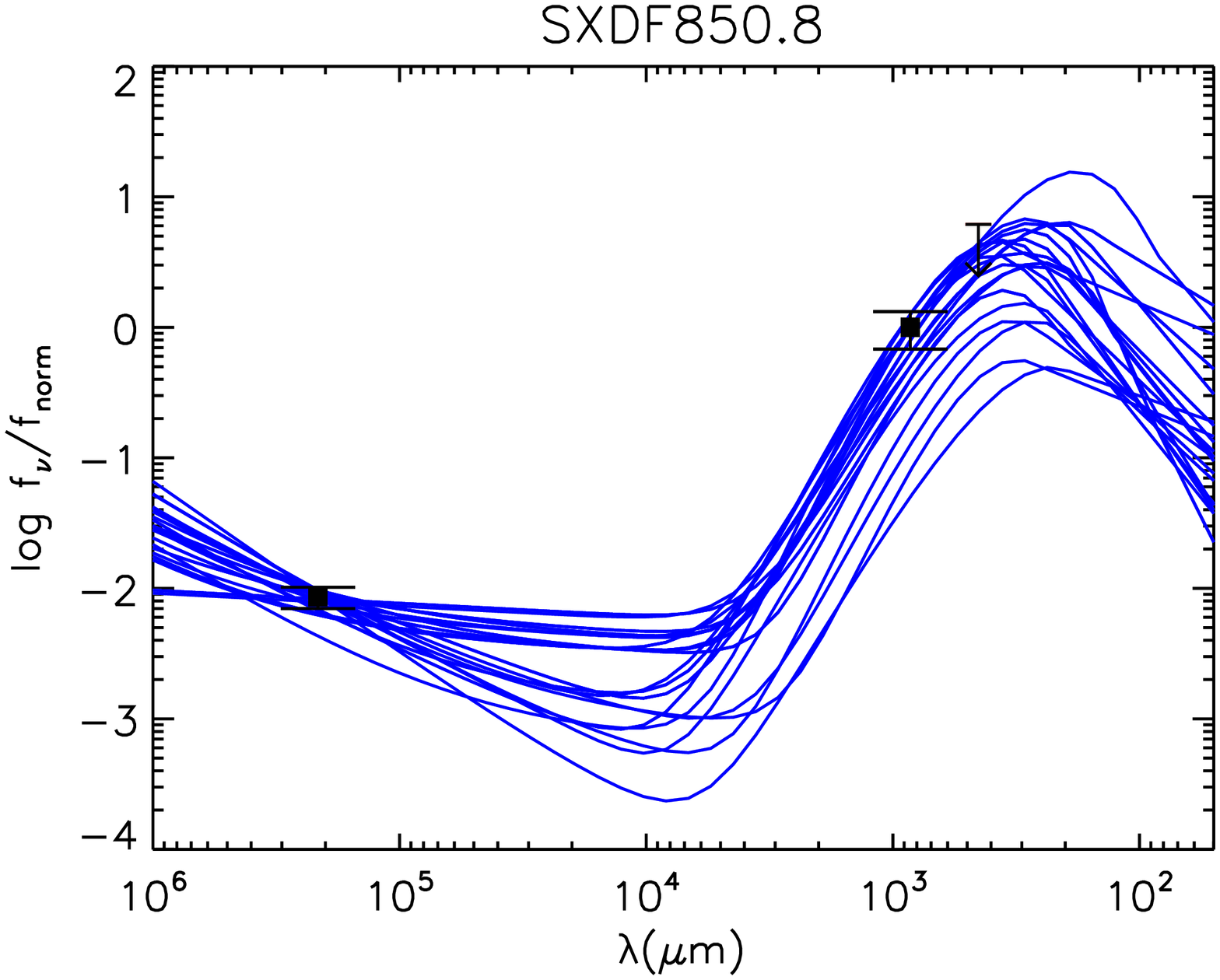,width=0.85\hsize,angle=90}
\hspace*{-1.1cm} 
\epsfig{file=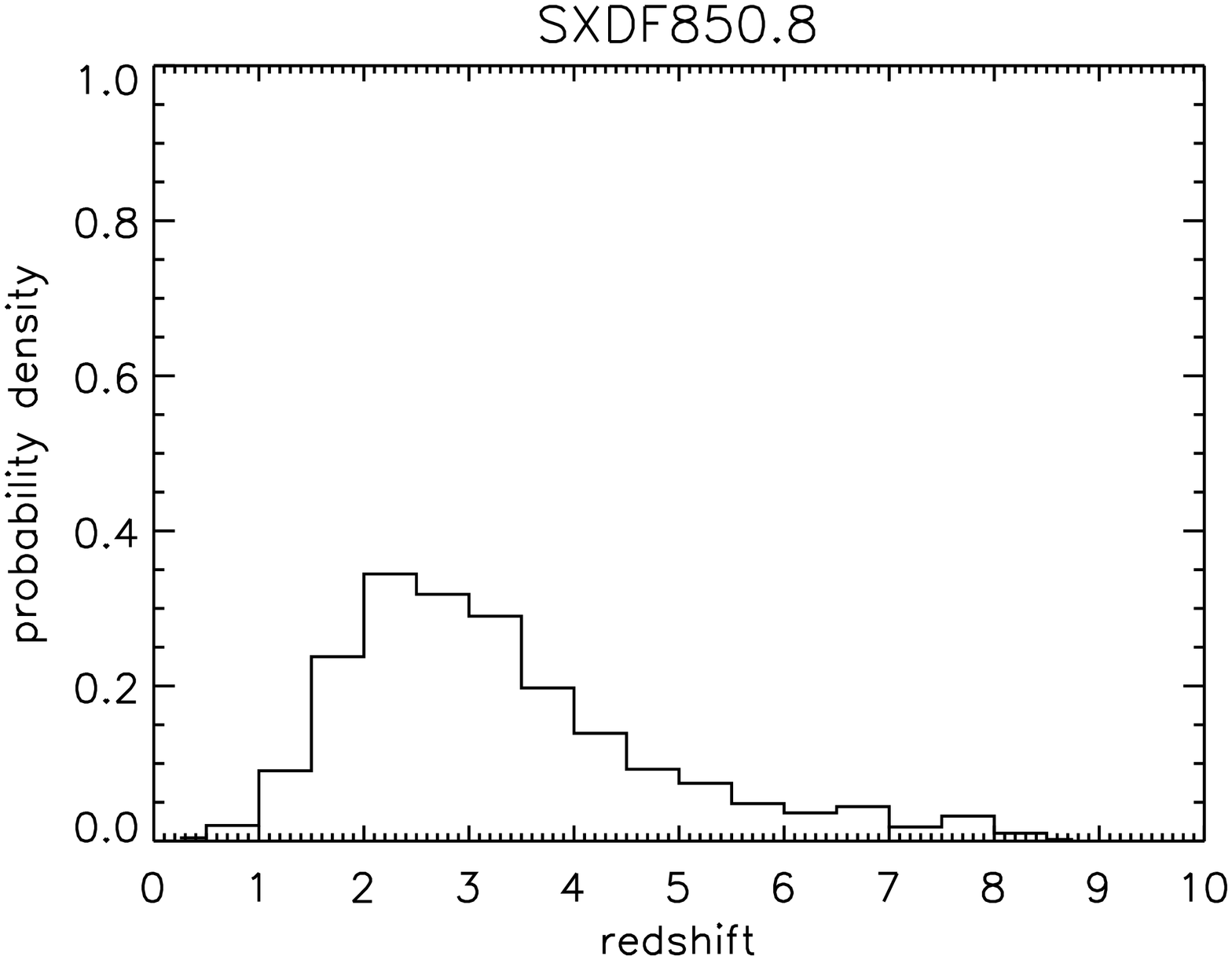,width=0.85\hsize,angle=90} 
%\end{figure}
%\begin{figure}
  \caption{ 
    {\bf (Top)} Spectral energy distribution (SED) of SXDF850.8, where 
the black squares mark the detections at 850$\mu$m and 1.4GHz used in
the photometric redshift calculation. Error bars are 1$\sigma$, and
the arrow at 450$\mu$m marks the 3$\sigma$ upper limit derived from
our maps.  For reference, the SED templates used in the
photometric redshift calculation are shifted to $z_{\rm phot}^{\rm A}=2.6$ and scaled to
maximize the likelihood function of detections and upper limit through
survival analysis (Isobe, Feigelson \& Nelson 1986), and 
are represented as lines. All
the SEDs are compatible within the 3$\sigma$ error-bars of the
photometry of the source.
  {\bf (Bottom)} Probability distribution for SXDF850.8 derived
for the $z_{\rm phot}^{\rm A}$ solution, 
using only the 850$\mu$m and 1.4GHz photometry.
 }
\label{fig:zexampleCY}
\end{figure}

\subsubsection{Redshift distribution of the SHADES population}

 We have assembled the best estimates of photometric redshift 
for each source in the LH and SXDF SHADES fields (figure~3). 
These are identified
by the
modes of the individual probability distributions, $z^{\rm A}_{\rm phot}$, which,
by definition, are the redshifts with the highest probability values.

\begin{figure}
\hspace*{-0.7cm} 
\epsfig{file=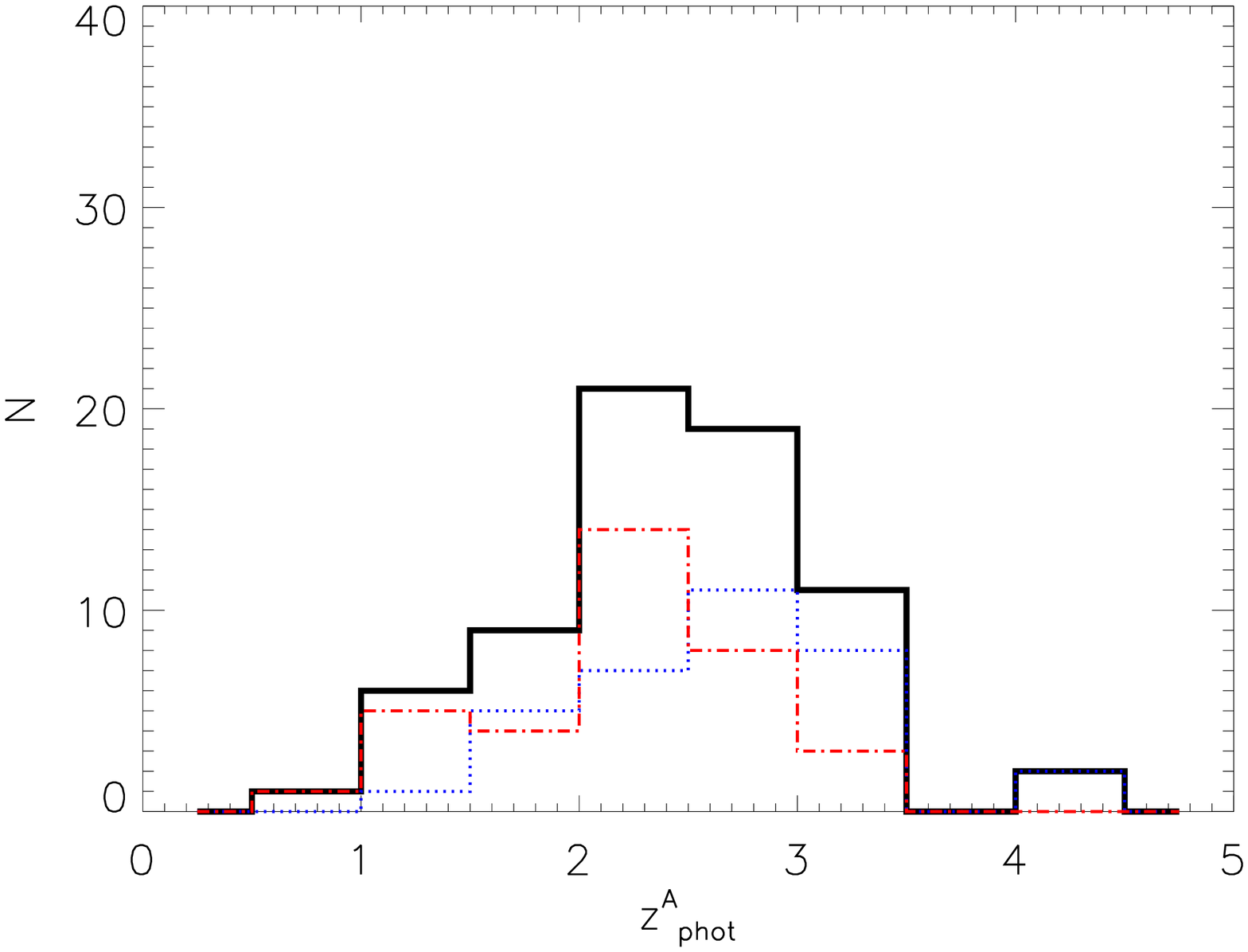,width=0.85\hsize,angle=90} 
\hspace*{-0.7cm} 
\epsfig{file=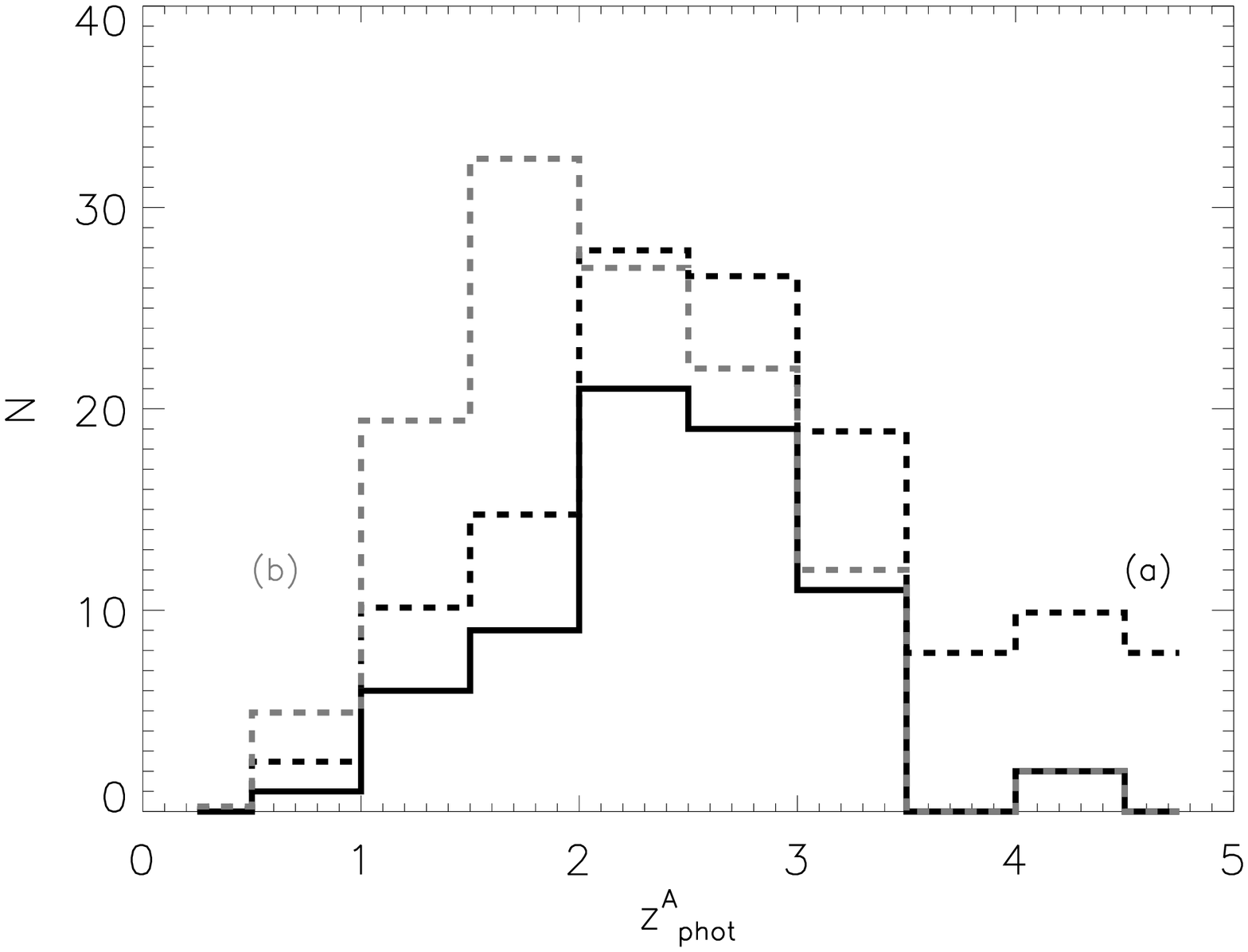,width=0.85\hsize,angle=90} 
\caption{ Histogram of modes of the photometric redshift distributions
of SHADES galaxies derived from the 1.4GHz/850$\mu$m spectral
index. The (black) thick solid-line (shown in the upper and lower-panels) represents 
the distribution of modes for the 69 galaxies that have been detected at 
both 850$\mu$m and 1.4GHz. In the upper-panel the (blue) thin dotted-line 
and (red) thin dash-dotted line represent the redshift distributions
in the LH  and SXDF fields respectively. In the lower-panel the black
dashed-line (a) and grey dashed-line (b) show the redshift
distributions for the full SHADES catalogue, including the 51 sub-mm galaxies
undetected at 1.4GHz. Those SCUBA galaxies with non-detections in the radio 
are distributed in one 
of two ways that bracket the range of reasonable options: (a)
with equal
probability between their calculated lower 90\% confidence limits 
and $z=5$; and alternatively, (b)  
between their lower limits and $z=2$, 
or only at their lower redshift-limits in the cases that these lie at $z>2$.}
\label{fig:zcumCY_A}
\end{figure}

The galaxies that are not detected with confidence at
radio-wavelengths, {\it i.e.} 26 out of the 60 galaxies in the LH, and
25 out of the 60 in the SXDF, have very flat individual redshift
probability distributions in our computations, and hence we quote only
the lower-limits to their redshift, which are defined as their 90\%
lower confidence-limits (tables 1 and 2). We incorporate these objects into the
population distribution by 
adding flat probability distributions
between their calculated lower 90\% confidence limits 
and $z=5$, and alternatively between their calculated lower 90\% 
confidence limits and $z=2$, or only at their lower redshift-limits 
if these indicate $z>2$. These two alternative 
priors illustrate how the resulting redshift distributions 
(that include SHADES galaxies without radio detections) 
can be biased high and low.

Figure~\ref{fig:zcumCY_A} shows the
final photometric redshift distribution using the 1.4GHz/850$\mu$m
spectral index, both for the full SHADES sample and for the LH and
SXDF fields separately.
The redshift distribution of SXDF sources peaks at slightly lower
redshifts (median $z\approx2.2$) compared to the distribution of LH
sources (median $z\approx 2.6$). The low-redshift tail ($z<1.5$) is
also slightly more prominent in the SXDF than in the LH (6 vs. 1
sources among the radio-detected sample). The difference in shape of
the two distributions of radio-detected sources can be measured using
the two-tailed Kolmogorov-Smirnov (K-S) test, which gives a 3~per-cent
chance that they are drawn from the same parent distribution. Furthermore,
the mean-redshifts of the two distributions are significantly
different at the 99.7~per cent level, according to a Mann-Whitney
$U$-test. This difference can be attributed to differences in the noise 
levels of the
radio maps, although some intrinsic variations between the fields 
are expected (see~\S~3).

For those objects with more than one redshift estimate (due to the
ambiguity in their counterparts), we have produced alternative
population distributions, with or without their inclusion, and with
different combinations of possible counterparts. The results do not
significantly change the final combined distribution. All figures
presented in this paper include the primary radio association if there
are multiple options.

\subsection{Radio-mm-FIR  SED analysis}

The SHADES fields have been targeted by other FIR/sub-mm/mm and radio
surveys. In this subsection we describe the 
extra constraints on the photometric redshifts that can be derived 
from these additional complementary data for a few tens of SHADES sources.

\subsubsection{Techniques and accuracies}

Photometric redshifts with modest precisions  ($\Delta z
\approx\ 0.3$ to 0.5) have been obtained in the
past few years using a combination of
spectral indices between the
radio and mm-wavelength regimes and the FIR spectral peak. 
This information has
been exploited by several groups using a wide array of
fitting-techniques and SEDs (e.g. Yun \& Carilli 2002, Hughes et
al. 2002, Aretxaga et al. 2003, 2005, Wiklind 2003, Hunt \& Maiolino
2005, Laurent et al. 2006). There remain, however, degeneracies
imposed by the choice of multiple SED templates, with FIR emission
peaks distributed over a range of wavelengths, which can limit the
precision of the derived redshifts (e.g. Blain, Barnard \& Chapman
2003).  We have previously developed a radio--mm--FIR technique based
on Monte-Carlo simulations, that take into account constraining prior
information such as the number counts of sub-mm galaxies, the favoured
luminosity/density evolution up to $z\approx 2$, and the lensing
amplification of a certain field (Hughes et al. 2002, Aretxaga et
al. 2003, 2005).  We only offer a brief summary of this technique
here.  A catalogue of 60$\mu$m luminosities and redshifts for mock
galaxies is generated from an evolutionary model for the $60\mu$m
luminosity function 
that fits the observed 850$\mu$m number-counts
(e.g. luminosity evolution $\propto (1+z)^3$ for $z\lsim 2$, and 
no evolution at $z>2$) and covers a simulated
area of 10~sq. deg.
Template SEDs are drawn at random, without regard
to their intrinsic luminosity, from a library of 20 local starbursts,
ULIRGs and AGN, to provide FIR--radio colours for the mock galaxies.
The SEDs cover a wide-range of FIR luminosities ($9.0 < {\rm log}
L_{\rm FIR}/L_{\odot} < 12.3$) and temperatures ($25<T/K<65$). The
flux densities of the mock galaxies include both photometric and calibration
errors, consistent with the quality of the observational data for each
sub-mm galaxy detected in a particular survey. We reject from the
catalogue those mock galaxies that do not respect the detection
thresholds and upper-limits of the particular sub-mm galaxy under
analysis.  The redshift probability distribution of an individual
sub-mm galaxy is then calculated as the normalized distribution of the
redshifts of the mock galaxies in the reduced catalogue, weighted by the
likelihood of identifying the colours and flux densities of each mock galaxy
with those of the sub-mm galaxy in question. This technique will be
denoted $z_{\rm phot}^{\rm MC}$ in the discussion that follows.

\begin{figure}
\hspace*{-1.1cm}
\epsfig{file=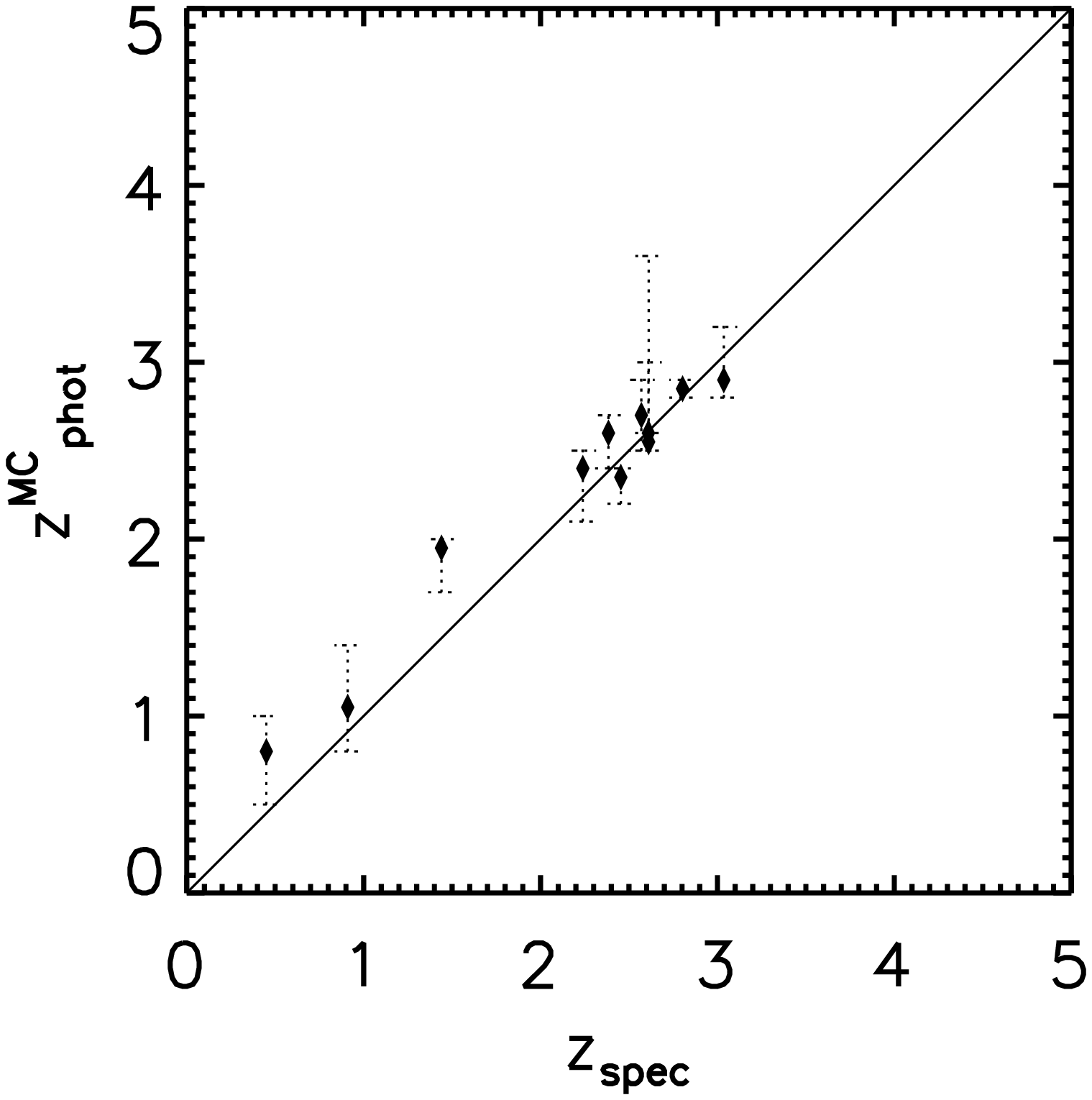,width=1.1\hsize,angle=90} 
\hspace*{-0.7cm} 
\epsfig{file=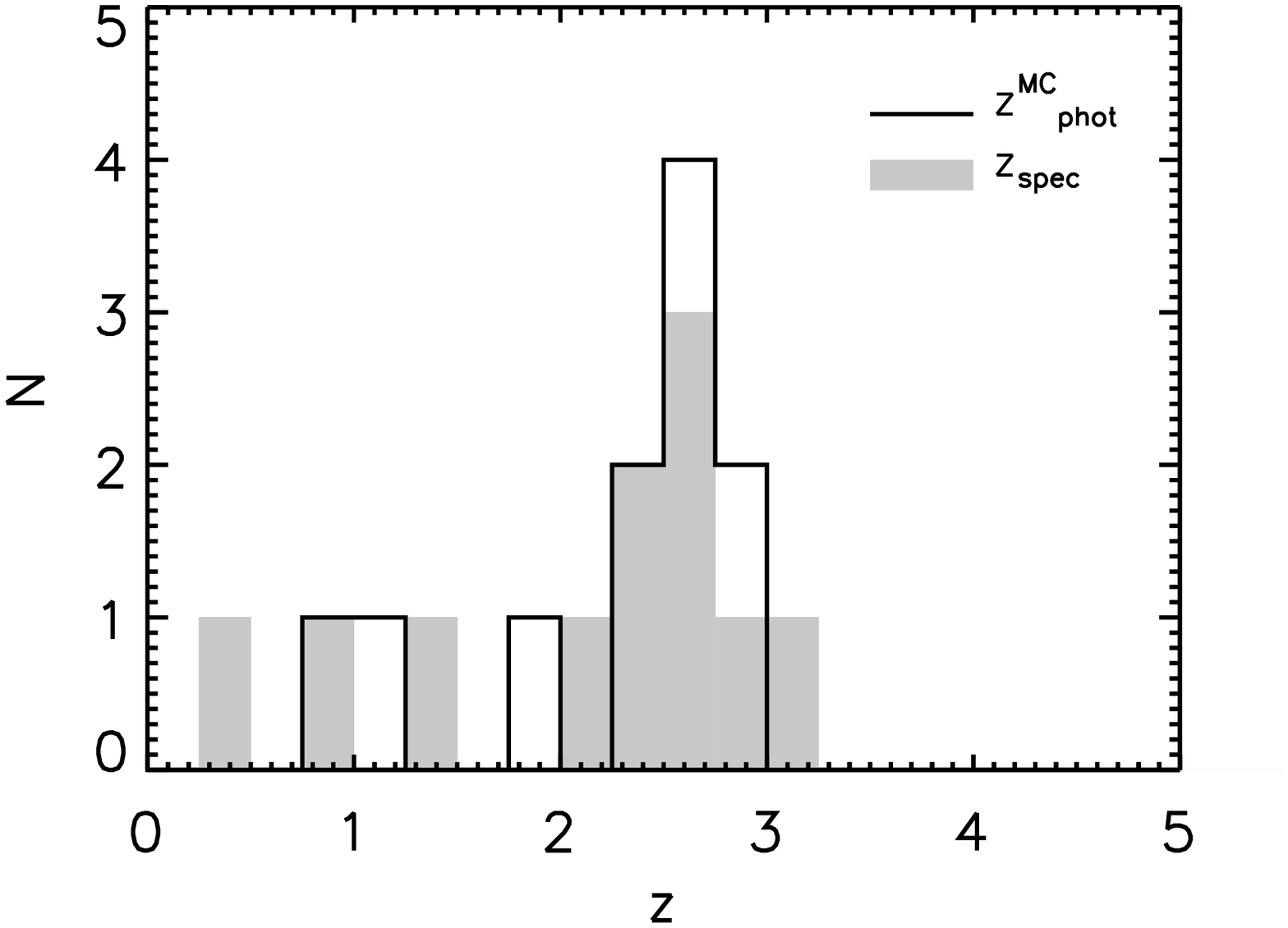,width=0.85\hsize,angle=90}
\caption{  {\bf (Top)} Comparison of spectroscopic and photometric redshifts
derived from the full radio-FIR SED for a sample of 11 sub-mm galaxies with
at least 3 robust detections at different wavelengths. This sample has
undisputed radio/optical/IR counterparts associated with the sub-mm
galaxies, and spectroscopic redshifts derived from 2 or more lines. The 
relationship has an r.m.s. of 0.25.
 {\bf (Bottom)} 
Comparison of the distributions of the
    spectroscopic and photometric redshifts represented in the top
    panel.  
}
\label{fig:zcum_A}
\end{figure}

   The validity of the results
  derived from this technique is limited by the assumption that the
  SEDs of high-$z$ sub-mm galaxies are similar to the local analogues,
  adopted as templates, which are scaled in luminosity and
  shifted in redshift. While this might seem a naive approach, all the templates
  used in the calculations that follow offer a good description of the
  radio--mm--FIR photometry of SCUBA galaxies, including 350$\mu$m 
  observations (Laurent et al. 2006, Kov\'acs et al
  2006), with known spectroscopic
  redshifts and un-ambiguous multi-wavelength counterparts (Aretxaga
  et al. 2005).  There are however a few sub-mm galaxies which do not match
  any of the templates we use in this paper at their published
  redshifts (see figure~4 in Aretxaga et al. 2006). In these examples
  their radio
  emission is higher than that implied by the radio--FIR correlation,
  possibly due to accretion activity, or their FIR emission
  peaks at wavelengths longer than those of the templates used in this
  study at the adopted redshift.  We describe the redshift solutions
  for these galaxies as `catastrophic' and this might be 
indicative of incompleteness in the library of SED templates used as analogues. 
There is
  still sufficient debate in the literature, however, about the 
nature and/or
  the ambiguity of the multi-wavelength counterparts to these sub-mm
  sources, from which the redshifts are derived, to justify their
  exclusion from a robust comparison sample (see Aretxaga et al. 2005,
  2006, Laurent et al. 2006, Kov\'acs et al. 2006 for a detailed
  complementary discussion on these galaxies).

In order to estimate the accuracy of the $z_{\rm phot}^{\rm MC}$
technique we use the full SED information of a robust sub-sample of 11
sub-mm galaxies, out of the comparison sample of 58 galaxies
considered in \S2.1.1, which have detections in three or more bands.
Furthermore, the same galaxies have undisputed identifications of their
optical/IR/radio counterparts and spectroscopic redshifts derived from
the measurement of two or more spectral lines. We derive a mean
accuracy for this sub-sample of $\Delta z \equiv <|z^{\rm MC}_{\rm
phot}-z_{\rm spec}|>\approx 0.2$ and an r.m.s. $< ( z^{\rm MC}_{\rm
phot}-z_{\rm spec})^2 >^{1/2} \approx 0.25$ over the whole redshift interval
(figure 4).  
Using all objects with
published photometry, regardless of whether the spectroscopic redshift
derived from the optical associations is ambiguous or not, the
overall accuracy over the $0\le z \le 4$ regime degrades to $\Delta z
\approx 0.55$, with an r.m.s. of 0.80 (see figure 3 in Aretxaga et al. 2006).  
A few significant outliers which remain in the correlation are 
discussed by Aretxaga et al. (2005) and  Kov\'acs et al. (2006). 
Within the small sub-sample of study, the accuracy is independent of redshift.

If we restrict the use of photometry to 450$\mu$m upper limits
combined with the 1.4GHz and 850$\mu$m detections for the comparison
sample of galaxies (adopting simulated 450$\mu$m upper limits, when
necessary, to mimic a shallower survey at this wavelength), we find a
mean accuracy of $\Delta z \approx 0.55$ and an r.m.s. of 0.7 for the
robust sample.  This result is especially relevant for the photometric
redshift calculations of the SHADES sources in \S\ref{sec:pzallSED},
the majority of which have similarly sparsely-sampled photometry.
Considering only the complete sample with 
robust and tentative spectroscopic redshifts the mean accuracy 
degrades slightly to $\Delta z \approx 0.65$, and the r.m.s. scatter 
is 0.90.

\subsubsection{The redshifts of SHADES sources}
\label{sec:pzallSED}

\begin{figure}
\hspace*{-1.1cm}
\epsfig{file=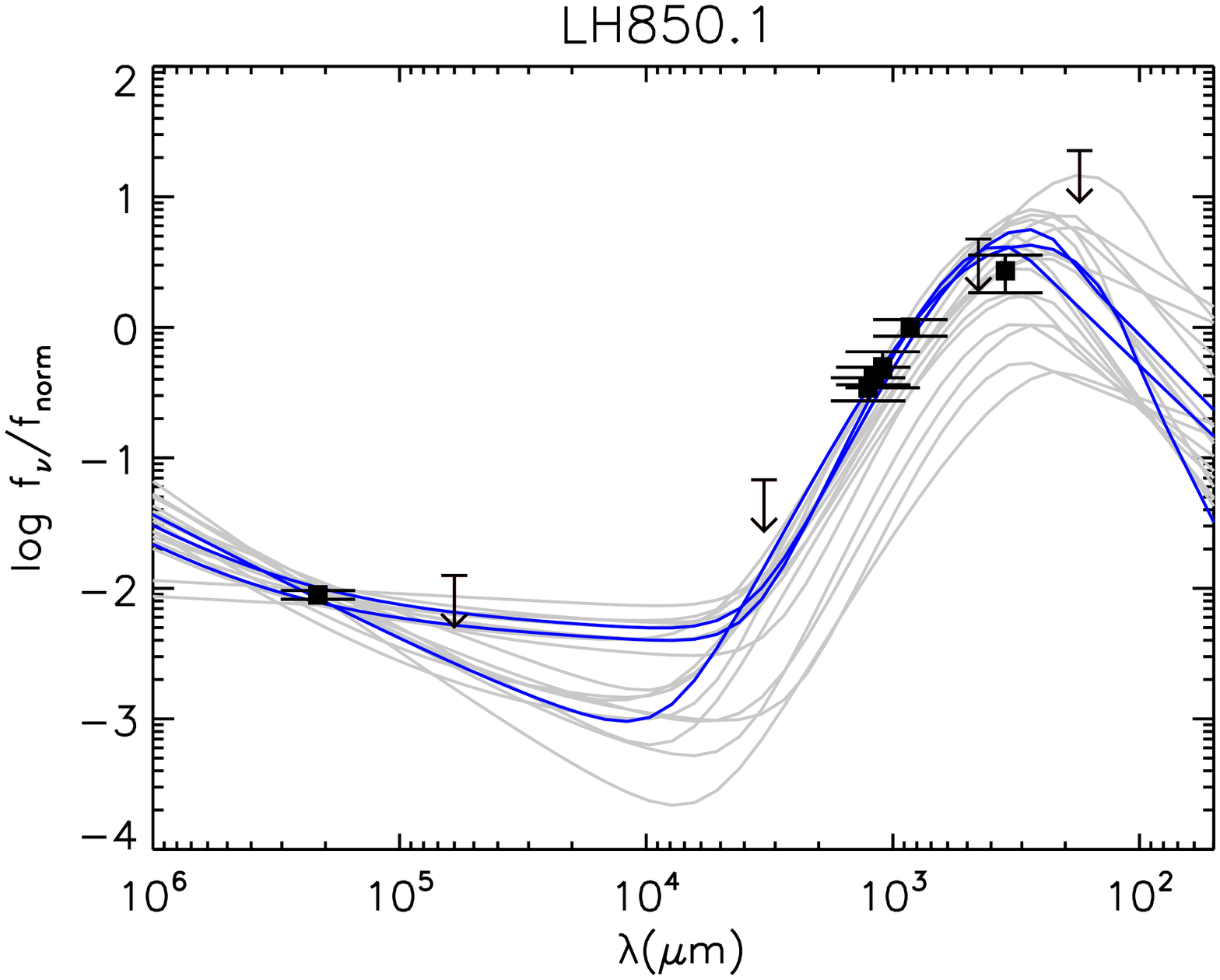,width=0.85\hsize,angle=90}
\hspace*{-1.1cm} 
\epsfig{file=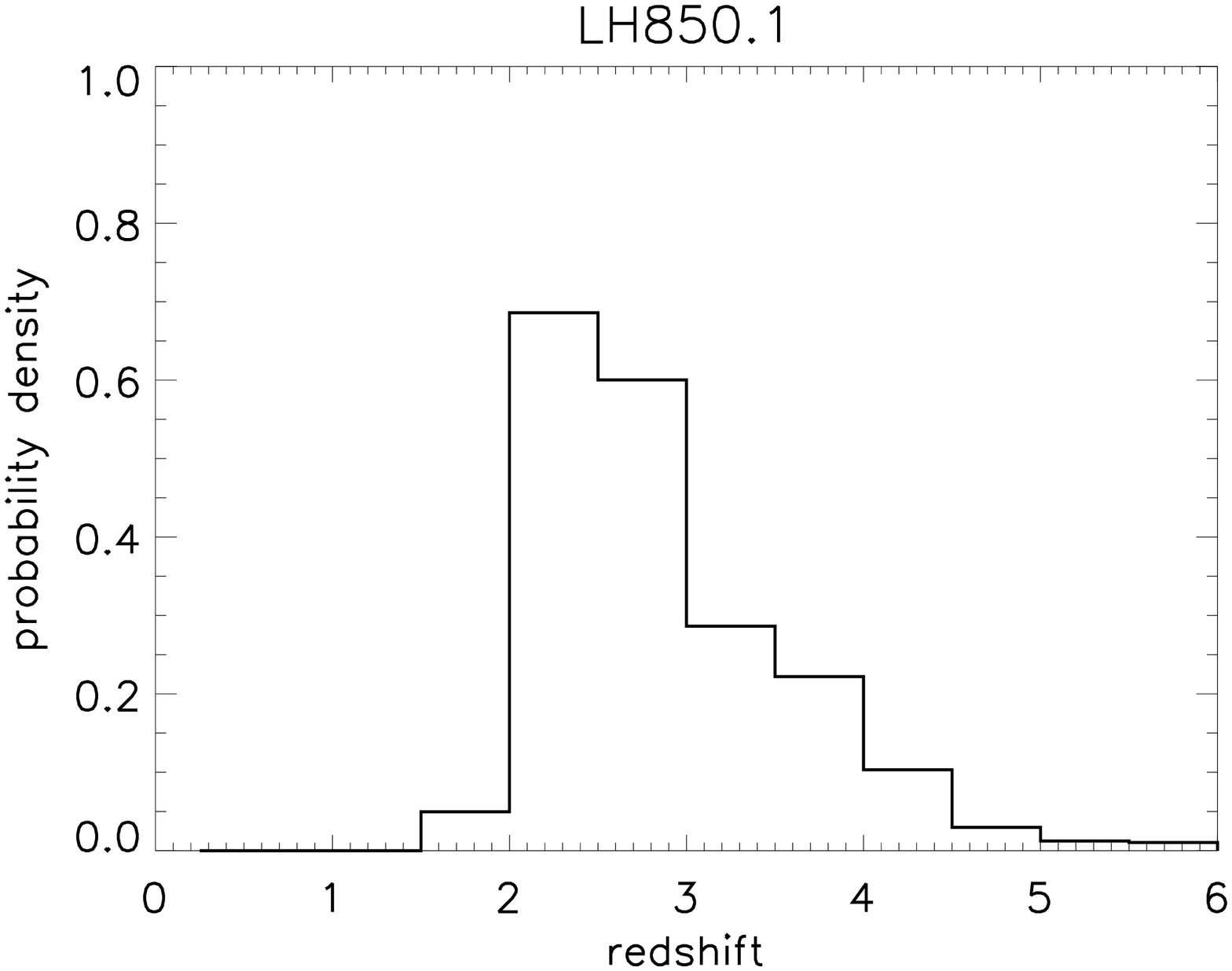,width=0.85\hsize,angle=90} 
%\end{figure}
%\begin{figure}
  \caption{ 
    {\bf (Top)} Spectral energy distribution (SED) of LH850.1, where 
the black squares mark detections, with 1$\sigma$ error bars, 
and arrows indicate 3$\sigma$ upper limits.
For reference, the SED templates used in the
photometric redshift calculation are shifted to $z^{\rm MC}_{\rm A}=2.4$ and 
scaled to
maximize the likelihood function of detections and upper limit through
survival analysis (Isobe et al. 1986), and are represented as lines. The 
SEDs compatible within the 3$\sigma$ error-bars of the
photometry of the source are represented in darker (blue) lines.
  {\bf (Bottom)} Probability distribution of LH850.1 derived
for the $z^{\rm MC}_{\rm A}$ solution.
 }
\label{fig:zexampleCY}
\end{figure}

Table~3 summarizes the most recent photometric redshifts calculated
with the Monte Carlo technique for SHADES sources with additional
photometry published in the literature.  In contrast, and for
completeness, tables~4 and 5 list the photometric redshifts derived
only from the combination of the SHADES 450/850$\mu$m and the 1.4GHz
photometry using two approaches: the Monte Carlo based technique,
$z_{\rm phot}^{\rm MC}$ described above, and a non-prior maximum
likelihood fit to the same 20 SEDs used for the first method that
includes survival analysis (Isobe, Feigelson \& Nelson 1986) to
incorporate the non-detections into the maximum likelihood formalism,
$z_{\rm phot}^{\rm SA}$.  This second technique is introduced to
provide a comparison of how the priors affect the redshift estimation
of the sources and the final combined redshift distribution of SHADES
galaxies. While most of the photometric redshifts derived from the two methods
are similar, the
pure survival analysis produces a few high-$z$ catastrophic results in
the robust comparison sample (2 out of 11). The overall reliability of
the maximum likelihood technique is $\Delta z \approx 0.7$. These high-$z$
catastrophic solutions get suppressed by the MC technique due to the weighting
priors that disfavour high-$z$ solutions for these sources, since,
if they were typical of the sub-mm population, they would overproduce
the 850$\mu$m number counts under the assumed luminosity evolution
model.  Although we give the values of photometric redshifts with and
without priors in tables 4 and 5, we will now continue the analysis of the
complete SHADES sample using only the MC solutions, since 
they have been shown to
perform better against the comparison sample.

For 5 sources in table\,5, SXDF850.5, 21, 28, 77 and 119, we also include
complementary photometry at 70 and 160$\mu$m from the Spitzer Legacy
Survey SWIRE (Lonsdale et al. 2003, Surance et al. 2007) that are used
to derive mid-IR counterparts to the SHADES sources (Clements et
al. 2007).  The remainder of the SHADES sources are not significantly
detected ($\ge 4\sigma$) in the Spitzer catalogues, and the noise
properties of the SWIRE maps, providing $3\sigma$ upper-limits of
$\sim$23~mJy and 160~mJy at 70 and 160$\mu$m, respectively
(Afonso-Luis et al. 2007), do not further constrain the
photometric redshifts.
 
Figure~5 shows the single-peaked redshift probability distribution 
derived for one of the sources that have the most complete photometric data. 
Although the majority of the sources 
show similar probability distributions, there are sometimes secondary peaks 
(see examples in Aretxaga et al. 2003).
Nevertheless, it is always the primary redshift-peak that defines the 
solutions given in tables 3,4,5.

\subsubsection{Redshift distribution of the SHADES population}

\begin{figure*}
\hspace*{-0.5cm}
\epsfig{file=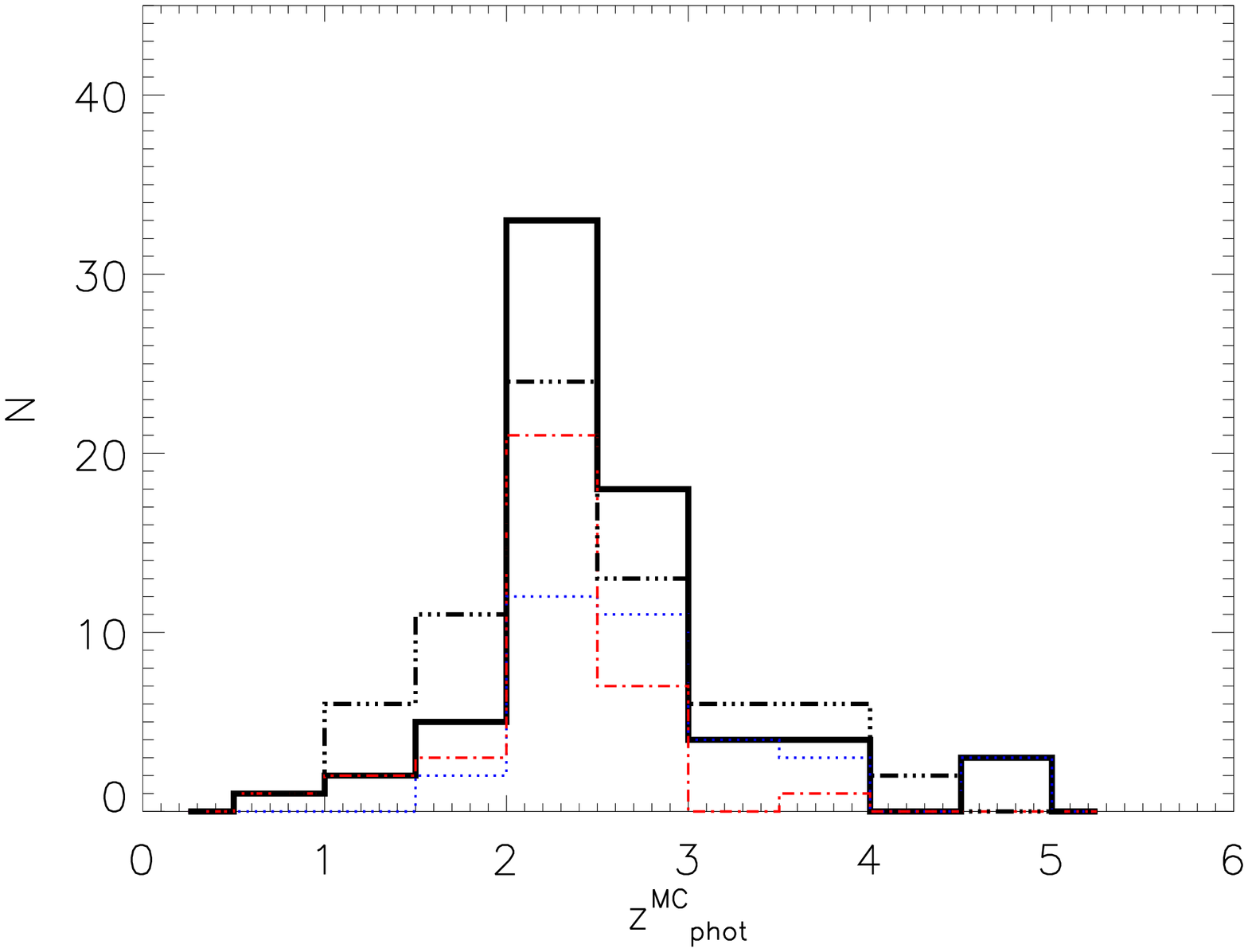,width=0.5\hsize,angle=90} 
\hspace*{-0.5cm}
\epsfig{file=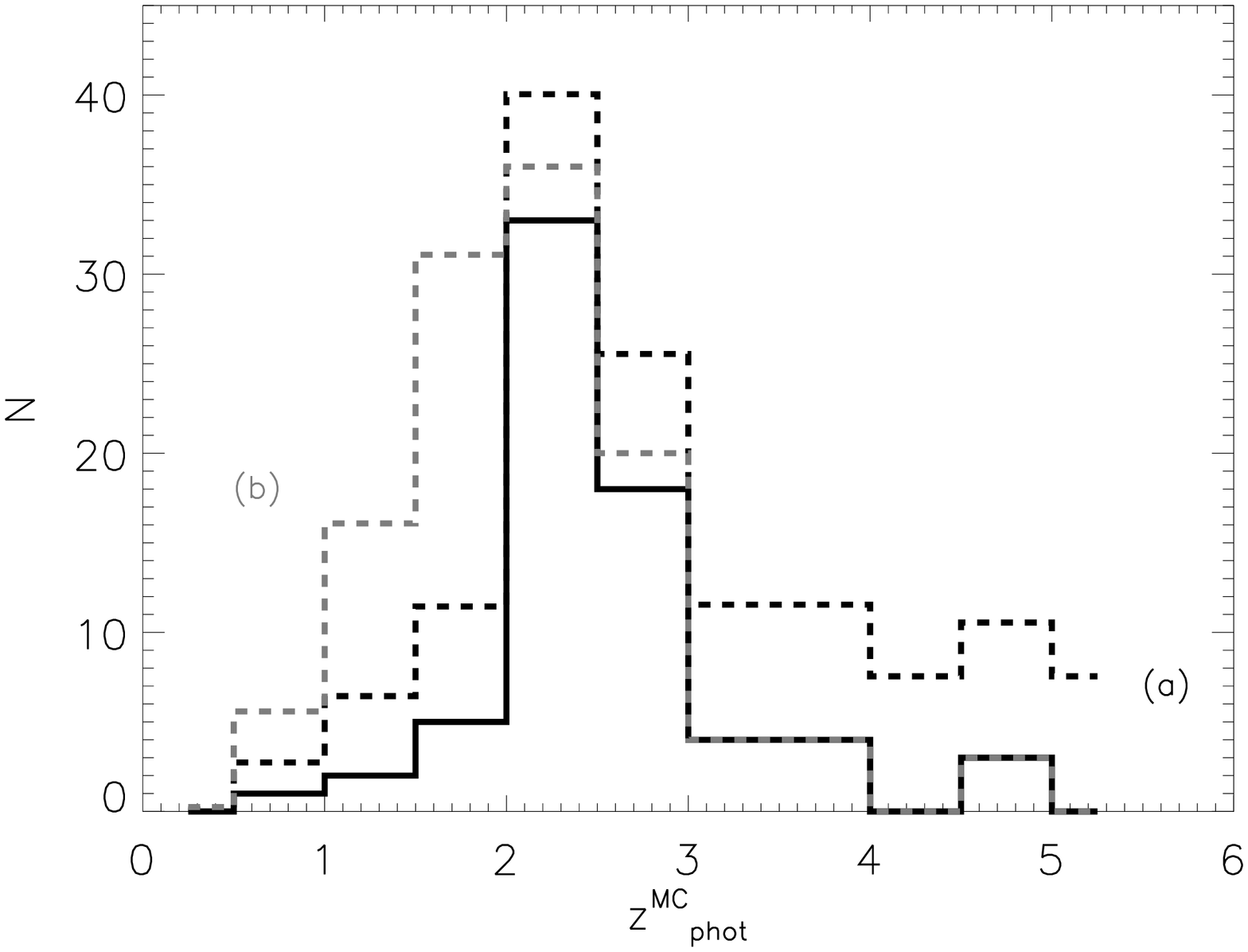,width=0.5\hsize,angle=90} 
\caption{ Histogram of the modes of the photometric
   redshifts of SHADES galaxies based on all available radio-mm-FIR
   photometry (provided in table 3, or otherwise in tables 4 and 5). 
   {\bf Upper Panel: } 
 The thick solid-line (also shown in the
lower-panel) represents the $z_{\rm phot}^{\rm MC}$ distribution of modes
   for those 70 SCUBA galaxies that have been detected in at least two bands,
   of which the thin (blue) dotted-line and thin (red) dash-dotted
   line represent the distributions defined by the LH and SXDF
   sources, respectively.   For comparison, we plot the
   $z_{\rm phot}^{\rm SA}$ solutions for the full distribution of 
   LH and SXDF (dash--3-dot line). 
{\bf Lower Panel: } The black dashed-line (a) and grey
dashed-line (b) show the redshift distributions for the 70
SCUBA galaxies detected in at least two bands, plus an additional 50 SCUBA
galaxies detected only at 850$\mu$m. 
These latter sources
are distributed in one 
of two ways that bracket the range of reasonable options: (a)
with equal
probability between their calculated lower 90\% confidence limits 
and $z=5$; and alternatively, (b)  
between their lower limits and $z=2$, 
or only at their lower redshift-limits in the cases that these lie at $z>2$.
}
\label{fig:zcum_MC}
\end{figure*}

Figure 6 shows our final photometrically-derived redshift distribution for the
SHADES sources, using our best available estimate for the redshift
of each source (i.e. $z_{\rm phot}^{\rm MC}$ taken from table 3 for those sources
with the most complete photometry, and from tables 4 and 5 for the remainder)

 The distribution of
radio-identified SHADES sources clearly peaks in the bin $z\approx 2.0-2.5$, 
with
a 50~per cent interquartile interval $z \sim 1.8 - 3.1$. 
We incorporate the non-radio detected sources (lower panel in figure~6)
into the
population distribution in two alternative ways, to serve as
examples of how much these sources could alter the final population
distribution: (a) approximating their individual probability
distributions as flat distributions between their lower 90\% confidence
limit and $z=5$; and (b) as flat 
distributions between their lower 90\% confidence limit and
$z=2$, or at their lower limit if this lies at $z>2$. Solution (a) is
actually derived from the adopted non-informative (flat) prior for the
photometric-redshift calculations. This creates a high-$z$ tail which
is a reflection of the adopted range for the flat redshift distributions 
which are unconstrained by the photometry.
Solution (b) is biased against high-$z$, 
by imposing a maximum redshift for the radio-undetected sample 
which is lower than the redshift of the peak of the radio-detected sample. 
 This radio-undetected sample could be composed of colder sub-mm galaxies
than those found in the template library, or they could have
the same template shapes as those adopted in the
photometric redshift analysis, and still be 
undetected at the depth of the present radio surveys.  Regardless, in these
alternative solutions, the mode of the population remains at
$z\approx 2.0-2.5$, with at least 50~per cent of the galaxies in the
interquartile range $1.6 \le z \le 3.4$.

In order to consider the effect of the objects with more than one
redshift estimate (due to ambiguity in their radio-counterparts), we
have produced alternative population distributions. For instance,
figure~6 shows the combination of the first entries for each source in
tables~3, 4, 5. The introduction of the second tabulated values
instead of the first ones, for those sources with ambiguous associated
photometry, produces an alternative distribution which is
indistinguishable (with a 99.96~per-cent probability), via a
Kolmogorov-Smirnov (K-S) test, from the one represented here.

As in the case of the photometric redshift distribution derived from
the 1.4GHz/850$\mu$m spectral index, the distribution of redshifts
derived from the full SED analysis of SXDF sources peaks at slightly
lower redshifts (median $z\approx2.2$) than that of LH sources
(median $z\approx2.7$), and its low-redshift tail ($z<1.5$) is also 
more prominent. These differences in the distributions are statistically-significant,
as indicated by a K-S test at a level of 98.9\%. A Mann-Whitney U-test 
shows that their mean-redshifts differ at the 99.997\% level.

\section{Discussion}

\subsection{Redshift distribution}

SHADES was designed with the objective of constraining the redshift
distribution and clustering properties of the sub-mm galaxy
population, an exercise which
van Kampen et al. (2005) demonstrated could discriminate between galaxy
formation models.  With  $\sim$40~per cent of the survey completed
before SCUBA was de-commissioned in the summer 2005, SHADES has
provided 120 robust sources. The radio-mm-FIR photometry assembled for
the survey favours a nearly Gaussian redshift distribution 
of the population peaking
at $z\approx 2.0-2.5$, albeit still with the possibility of a 
high-redshift tail remaining.

The photometric redshift distribution of the radio-detected sub-mm
galaxies is qualitatively similar to the optical spectroscopic
redshift distribution published by Chapman et al. (2003, 2005) who
followed-up a sample of sub-mm galaxies derived from various surveys.
The agreement is perhaps not surprising, given that the photometric
redshifts of the comparison sample have shown a relatively good
agreement with the spectroscopic redshifts published in the literature
(Aretxaga et al. 2006, \S2.1.1, 2.2.1). Furthermore, the majority of
the sub-mm sources that have spectroscopic redshifts are drawn from
SCUBA surveys of similar depths to SHADES.  

The high-redshift correction applied to the measured spectroscopic
redshift distribution, suggested by Chapman et al. (2005) to account
for the bias introduced by non-detection of the higher-redshift radio
counterparts that provide candidates for optical spectroscopic follow-up, also
falls within the range of photometric redshift estimations we have
derived for SHADES sources that are not detected at radio-wavelengths.
 These
sources provide the high-redshift ($z > 3$) tail of figure~6, and could
in fact be placed anywhere above $z\sim 1.0$, even producing secondary peaks.

Our calculations do not support the existence of a substantial
low-redshift ($z < 1.5$) tail within the luminous sub-mm 
radio-detected population
sampled by SHADES. At first sight this might appear to be in conflict with the
results of Pope et al. (2005, 2006) who found that $\sim 30$\% of the sub-mm
sources found in the SCUBA imaging of the GOODS-North field may lie at 
$z < 1.5$. If the GOODS-North 850$\mu$m catalogue is restricted to sources 
with de-boosted flux densities $S_{850} > 3$mJy, however, then the
proportion of robustly identified sub-mm galaxies which lie at $z <
1.5$ drops to 6\%. This is entirely consistent with the results found here for
SHADES galaxies. Thus, these results may be providing
further evidence that the peak of the redshift distribution of sub-mm sources
is positively correlated with sub-mm flux-density/luminosity, consistent
with the apparently anti-hierarchical nature of star-formation history
reported in several other recent studies (e.g. Heavens et al. 2004).
Furthermore field-to-field variations in the spatial distribution of the 
large-scale structure can provide a simple explanation for the 
differences between the redshift distributions of sub-mm sources derived 
from the individually-mapped contiguous-areas (typically $< 0.2$\, sq. 
degrees) taken from the current generation of SCUBA surveys.

Our analysis also suggests that only a modest fraction of sub-mm galaxies 
could be hiding in the optical redshift-desert at $z\approx
1.5-1.8$ during spectroscopic searches for SHADES sources with robust radio counterparts.
The photometric redshift probability density distributions 
of radio-detected SHADES sources
using the 1.4GHz/850$\mu$m index or the full radio--mm--FIR SED
information contain $\sim 15$~\% and $\sim 10\%$ of sources in
this redshift desert regime, respectively.

  The difference between the redshift distribution of sub-mm sources
in the SXDF and LH fields is entirely consistent with the different properties
of the 1.4-GHz maps as discussed by Ivison et al. (2007). The LH data
have a higher VLA resolution than those in SXDF, and the LH data are also
deeper, although the coverage is less uniform. There is clearly
potential for systematic differences between radio measurements in the
LH and SXDF. For an extended source in the LH (of which there are
several -- Ivison et al. 2002), a larger fraction of emission on
scales larger than the synthesized beam will be resolved away than for
similar cases in SXDF. Moreover, the LH data will suffer greater
significant bandwidth smearing and, although the appropriate
correction has been made to the measured flux densities, some faint
sources will be lost below the radio-detection threshold and may receive
misleadingly low flux-density limits.  These effects can be viewed as
a systematic flux calibration offset with consequences as severe as
those encountered in optical/infrared photometric-redshift estimation.
While random 1.4GHz calibration uncertainties of 5~per cent have been
accounted for in the estimation of the photometric redshifts, a
systematic flux-density offset could shift the redshift distribution
significantly. In order to explore this possibility, 
we have applied a 10~per cent flux increase to
the LH photometry and recalculated the photometric redshifts. The
combined redshift distribution shifts its peak by $\sim -0.25$, and
consequently the mean values of the SXDF and LH distributions are more
consistent, increasing from 0.3~per cent (\S 2.1.3) to a 7~per-cent
probability, according to a Mann-Whitney $U$-test. 
Some intrinsic variation on the distribution of
redshifts between the fields is however to be expected (see below).

\subsubsection{Comparison of the SHADES redshift distribution with  galaxy formation models}

\begin{figure*}
\hspace*{-0.7cm}
\epsfig{file=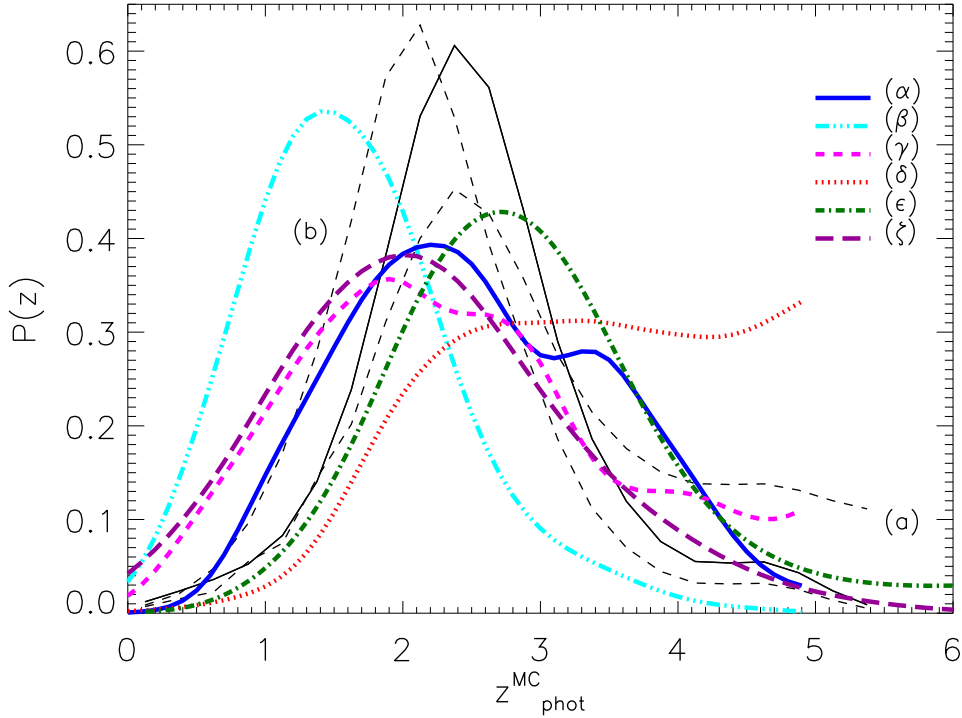,width=0.6\hsize,angle=90} 
\caption{ Probability density of the combined redshift distribution of
  SHADES galaxies (thin black solid-line, and thin dashed-lines ((a) and
  (b)), as described in figure~6). These are compared with the
  redshift distributions of six galaxy formation models (thick
  coloured-lines), degraded with a $\sigma_z =0.4$ to provide representative 
 redshift-uncertainties: ($\alpha$) the hydrodynamical model
  of Muanwong et al. (2002) coupled with the analytical form for
  redshift distribution of Baugh, Cole \& Frenk (1996); ($\beta$) the
  simple merger model of van Kampen et al. (2005); ($\gamma$) the
  phenomenological model of van Kampen (2004) and van Kampen 
  et al. (2005); ($\delta$) the stable
  clustering model of Gazta\~naga \& Hughes (2001); ($\epsilon$) the
  semi-analytical model for the joint formation and evolution of
  spheroids and QSOs of Silva et al. (2005); and ($\zeta$) the
  semi-analytic model for galaxy formation of Baugh et al. (2005).  }
\label{fig:zcum_A}
\end{figure*}

Van Kampen et al. (2005) studied four different galaxy formation
models that yielded different redshift distributions and clustering
properties for the sub-mm population expected to be found in a survey
of the depth and area covered by SHADES: ($\alpha$) a
hydrodynamical model (Muanwong et al. 2002), that follows the
evolution of dark-matter, gas, star-like particles and galaxy
fragments, that has been coupled with the analytical form for redshift
distribution of Baugh, Cole \& Frenk (1996); ($\beta$) a simple merger
model that identifies sub-mm galaxies with major mergers of 
massive galaxies; ($\gamma$) a phenomenological model (van Kampen
2004), which is based on $N$-body simulations that identify the sites
of major-mergers and has two modes of star formation, quiescent and
bursting; and ($\delta$) a stable clustering model
(Gazta\~naga \& Hughes 2001).  Figure~7 represents the theoretical
redshift distributions of SHADES galaxies found in these models. This
figure has been complemented with ($\epsilon$) a semi-analytical model
for the joint formation and evolution of spheroids and QSOs (Granato
et al. 2004, Silva et al. 2005); and ($\zeta$) an alternative semi-analytic
model of galaxy formation for sub-mm galaxies (Baugh et al. 2005). 

Furthermore, to enable a more accurate discrimination between the
above predictions, all the galaxy formation models in figure\,7
account for the incompleteness of sources in the SHADES catalogue
(Coppin et al. 2006). The models have also been convolved with a
representative radio-mm-FIR photometric precision of $\sigma \sim 0.4$,
which is intermediate between the measured uncertainties derived for
the two techniques used in this paper.

We have made a comparison, via a K-S test, of the 
observed redshift probability density distributions with those 
predicted from the above 
models.  In each case a K-S statistic has been calculated that 
accommodates the 1$\sigma$ uncertainty in the median redshift of the models 
due to field-to-field variations. A study of 25 simulations made for
each of the four models analyzed by van Kampen et al. (2005) demonstrate 
that the mean redshift of $\sim 60$ SHADES-like sub-mm galaxies varies
by $\delta \bar{z}$ (r.m.s.) $\approx 0.25-0.55$. In part these shifts
can be explained by Poisson noise (estimated $\sigma \sim 0.1 - 0.2$
from the simulations).  The models show, however, that there
could also be a significant component in the field-to-field variations
that arises from intrinsic redshift differences due to varying amounts
of groups or proto-clusters of galaxies along the line-of sight. Thus
the differences found between the LH and SXDF areas, and between these
and smaller, deeper surveys like GOODS-N, could be partially explained
by this effect.

The results of the K-S test suggest that
only model\,($\epsilon$) is close to being formally acceptable, with 
an 87\% probability for the model to  agree with the measured probability density
distribution that includes SHADES sources with and without radio-detections
according to solution (a). With only a small shift ($\delta z \sim -0.3$) 
in the distribution, model\,($\epsilon$) also qualitatively reproduces 
($\sim 60\%$ probability of similarity) the photometric-redshift distribution 
of the radio-detected SHADES galaxies. 

The SHADES sources in our analysis that are not detected at radio wavelengths 
have very flat redshift probability distributions, which simply places them 
at $z\gsim 1.0$, and hence these SHADES sources could also produce a secondary 
peak in the redshift distribution.
In the ranking of similarities of measured and model-distributions, 
models\,($\alpha$) and ($\gamma$), $\sim 45$\% probability, have 
double peaks and are broader 
than the observed distributions. A different prior, that optimizes the redshift-distribution 
of the SHADES sources without radio-detections, could bring them closer
to a level of formal-acceptance. 
Finally models\,($\delta$), ($\zeta$) and ($\beta$) are all rejected with probabilities of 
$< 2\%$ of being consistent with the range of solutions depicted in figure\,7.

\subsection{The FIR luminosity of SHADES sources}

\begin{figure}
\hspace*{-0.7cm}
\epsfig{file=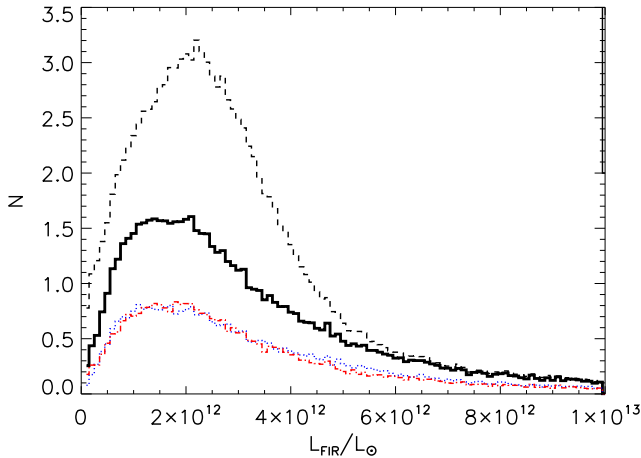,width=0.8\hsize,angle=90} 
\caption{Distribution of FIR luminosities of the 120 SCUBA galaxies in
the SHADES catalogue. The solid-line represents the total distribution
of radio-detected sub-mm sources in the two fields towards the LH
(thin blue dotted-line) and SXDF (thin red dashed-dotted line) based
on the FIR-radio correlation of sub-mm galaxies (Kov\'acs et
al. 2006). The dashed line shows the complete distribution of FIR
luminosities for all SHADES sources, including the 51
non-radio-detected sub-mm sources, where the FIR luminosities of the
latter sources have been derived from a single $T=35$K, $\beta=1.5$
grey-body scaled to the observed 850$\mu$m flux density, and their redshifts
have been selected at random between their lower 90\%-confidence
redshift-limits and, arbitrarily, $z=5$.}
\label{fig:lum}
\end{figure}

The catalogues of redshifts presented in tables~1 to 5 are an initial
step towards characterizing the FIR luminosities and star formation
rates of the SHADES population.  The available photometry in the FIR
peak regime (70--450$\mu$m), however, is not deep enough to fully constrain 
the SEDs of most SHADES sources at these wavelengths. One viable
approach is to use the 20 SEDs in our local template catalogue to
derive the corresponding FIR luminosities from the 850$\mu$m flux densities,
bearing in mind that the lack of constraints at short wavelengths will
dominate the errors in luminosity estimation over those of redshift
(e.g. Hughes et al. 2002).  Alternatively, one could use the 1.4GHz
radio flux density to deduce FIR luminosities via the radio-FIR luminosity
correlation that characterizes the sub-mm galaxy population, since this 
now has been extended to $z\sim 0.5-4$ (Kov\'acs et al. 2006). This
latter approach has the advantage of providing mean FIR luminosities which
are accurate for the bulk of the population, reducing the
uncertainties in luminosity primarily to the accuracy of the
photometric redshifts. However, the normalization of the relation might be
shifted from the local IRAS correlation, and this could affect the
FIR luminosities derived, and the comparison of these to nearby galaxies.

Regardless of this complication, the observed 1.4GHz flux densities have been
converted to rest-frame 1.4GHz flux densities using a mean synchrotron radio
slope of index $\alpha=-0.7$, and the monochromatic 1.4GHz luminosity
has been inferred using the photometric redshift solution for each
source. This is converted to FIR luminosity using the linear
relationship $\log (L_{\rm FIR}/ L_{1.4{\rm GHz}}/ 4.52\,{\rm
THz})=2.14\pm 0.07$ (Kov\'acs et al. 2006).  For each source we have
considered the effect of the uncertainties in redshift, 1.4GHz flux 
and the reported scatter in the sub-mm galaxy FIR-radio correlation
(Kov\'acs et al. 2006) by bootstrapping 1000 times on the measured
errors. For the 69 radio-detected SHADES sources in the LH and SXDF
fields the median FIR luminosity is $2.6\times10^{12}$\Lsun, with a
high luminosity tail that extends to $~1\times10^{13}$\Lsun\ (see
figure~\ref{fig:lum}). The distribution of luminosities for sources in
both the LH and SXDF fields are similar.

The effect of non-radio-detected sources, which have very
unconstrained and possibly high redshifts (\S2.1), has also been considered by
scaling a simple grey-body of temperature $T=35$~K and emissivity index 
$\beta=1.5$, the average of a parametrized SED of the short
sub-mm wavelength detected SCUBA galaxies that define the radio-FIR 
correlation at $z\sim 1-3$ (Kov\'acs et al. 2006), to
the observed 850$\mu$m flux density. The errors in the 850$\mu$m flux density 
have been
taken into account by bootstrapping on the inferred de-boosted
distributions of 850$\mu$m flux densities for each source, and the redshift
has been selected at random between their 90~per cent lower-limits
(tables~4, 5) and, arbitrarily, $z=5$. The resulting combined distribution has a median
FIR luminosity of $2.6\times10^{12}$\Lsun, the same as the
distribution of radio-detected sources. 
%The change in the redshift limits,
%from their 90~per cent lower limits to a fixed value of $z\sim 3$, for
%the non-radio detected sources does not significantly change the
%luminosity estimates, as the K-correction of the 850~$\mu$m flux density is
%essentially constant at these redshifts.

If one returns to the alternative approach of using the combination of 
20 SEDs with the de-boosted 850$\mu$m flux densities to derive the FIR luminosities, one derives a broader luminosity distribution than that depicted in figure~8, reflecting the wide variety of acceptable SED templates,  with a median 
luminosity that is increased by $\sim 40$~per cent.

\subsection{The star formation-rate history derived from SHADES galaxies}

\begin{figure*}
\hspace*{-0.7cm}
\epsfig{file=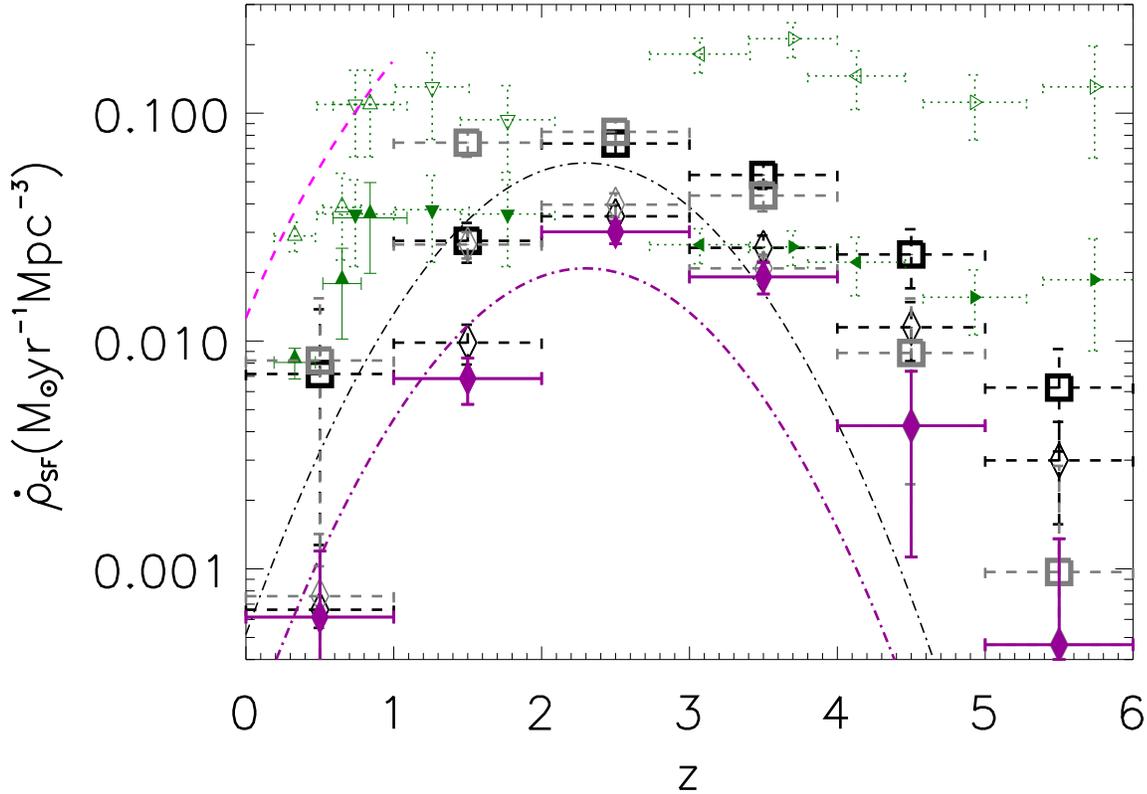,width=0.7\hsize,angle=90} 
\caption{Evolution of the global star formation rate density (SFRD)
  for different samples of galaxies. We show the SFRD for
  radio-detected SHADES sources (solid diamonds) and for all SHADES
  galaxies, considering both the radio-detected and radio-undetected
  sources (open diamonds).  The redshift probability distributions for
  the radio-undetected sources have been assumed to be flat between
  their lower redshift-limits (tables~4, 5) and $z=2$ (grey open
  diamonds) and $z=5$ (black open diamonds).  The redshift error-bars
  indicate the width of the redshift bins.  The error-bars in
  $\dot\rho_{\rm SF}$ are a combination of the uncertainty in the
  photometry, the recovery of the luminosity of each source due to SED
  differences, and the uncertainty in redshift that divides the
  contribution of sources across several redshift bins (see \S 3.3).
  The empty-square grey/black symbols represent the SFRD traced by ultraluminous
  $850\mu$m-selected starbursts, by correcting the SFRDs derived for
  SHADES galaxies via the completion of the IRAS 60$\mu$m luminosity
  function with pure luminosity 
  evolution up to $z=2$ (\S 3.3).  The (pink) dashed-line shows the SFRD inferred
  from 24$\mu$m-selected luminous and ultraluminous infrared galaxies
  (Le Floc'h et al. 2005), which follow a similar luminosity evolution
  to that presented here for the SHADES galaxies.  The (purple) thick
  dash-dotted line shows the SFRD for the sample of SCUBA galaxies
  from Chapman et al. (2005). The thin dash-dotted line, which is a
  factor of $\sim 3$ higher, is an estimation of the contribution of
  850$\mu$m-selected galaxies down to the $\sim 1$\,mJy level.  The
  SFRDs for optical/UV-selected starbursts are shown as small 
  triangles, and are taken from Lilly et al. (1996 -- upward-pointing
  triangles), Connolly et al. (1997 -- downward-pointing), Steidel et
  al. (1999 -- left-pointing) and Giavalisco et al. (2004 --
  right-pointing).  The optical/UV data are shown with and without
  corrections for dust-obscuration as empty and solid triangles
  respectively, and all data have been homogenized to the same set of
  parameters and corrected to complete a Schechter luminosity-function
  (Giavalisco et al. 2004).
All
estimates have been converted to the same SFR/$L_{\rm IR}$ factor and
cosmological model described in \S\,1.  
}
\label{fig:sfr}
\end{figure*}

The evolution of the global star formation rate (SFR) density traced
by SHADES sources is shown in figure~\ref{fig:sfr}. The conversion
from FIR luminosity to star formation rate was performed using a
constant of $5\times 10^9$~\Lsun /(\Msun yr$^{-1}$), which is
constrained to $\sim \pm 30$~per cent uncertainty (Kennicutt 1998).
We have multiplied the contribution of each source to the star
formation density by the inverse of the SHADES survey completeness
function at the appropriate flux density (Coppin et al. 2006).
Redshift-space was binned into six intervals, and a Monte Carlo was
performed to assign each galaxy to a redshift bin according to its
expected photometric redshift error.  The FIR luminosities have been
estimated from the radio-FIR correlation or with a single SED which is
considered to be representative of the sub-mm galaxy population (as in
\S3.2). The error bars in SFR density are the result of the
uncertainties in photometry, SEDs and redshift, and are computed as
the standard deviation traced by 1000 Monte Carlo simulations.

The evolution in the SFR density traced by the radio-detected sources
shows a clear peak at $z\sim 2.5$ and a slow decline at both low and
high redshifts.  A good description of the SFR density traced by this
population is given by $\dot{\rho}_{\rm SF} \approx 0.35 \exp
\left[-0.5 (z-2.8)^2/(0.8^2) \right] $.  The contribution of
non-radio-detected SHADES sources to the SFR history may be
significant at high-redshifts compared to the radio-detected SHADES sources,
as indicated in figure~9 by the empty black diamonds (solution (a) in
figure~6). Since the redshifts of these sources are not well
constrained, this should be considered as only a possible evolutionary
history, awaiting confirmation by better multi-wavelength data to
improve the constraints on the redshifts of the radio-undetected
SHADES sources.  The effect of placing all the radio-undetected
sources between their lower redshift-limits and $z=2$ or at their
lower limits if they are at $z>2$ (solution (b) in figure~6) is shown
by the grey empty diamonds in figure~9, which understandably broadens
the peak of star formation density to lower-$z$.

The levels of star formation deduced for SHADES sources are consistent
with those derived for other samples of radio-detected sub-mm galaxies
at $z\le 3.5$ (Chapman et al. 2005). At higher-redshifts, however, the
SFR densities implied in this photometric-redshift study exceed the
extrapolations of Chapman et al.  Eventually spectroscopic
measurements of SHADES sources at $z > 3.5$, via millimetre-wavelength
observations of molecular CO-lines, or optical spectroscopy with
increased sensitivity, will provide a definitive measurement of the SFR
density of obscured galaxies in the high-redshift Universe.

In order to estimate the contribution to the global star formation
rate density of $850\mu$m sources that are fainter than those detected
by SHADES, we have adopted the $60\mu$m luminosity function with a
pure luminosity evolution that follows $(1+z)^3$ at $z \le 2$, and then
maintains a constant level for $z > 2$. This evolutionary form
provides an adequate description of the 850$\mu$m and 1.1mm counts
(e.g. Scott et al. 2002, Greve et al. 2004) and is also supported up
to $z=1$ by the evolution of the luminosity function of
24$\mu$m--selected galaxies (Le Floc'h et al. 2005). By implementing this
luminosity-function correction the SFR density increases up to a maximum 
factor of 2 at $z>2$.

The contribution of SHADES galaxies to the global star formation rate
density of the Universe is comparable to the contribution of
starbursts selected at optical/UV wavelengths at $1\lsim z\lsim 4$
before the latter are corrected for dust extinction. It is important
to recall that the $L_{\rm FIR}/$SFR factor we have adopted could be
in error by $\sim \pm 30$~per cent, and that SED differences could
also account for an increase of $\sim 40$~per cent. These uncertainties
have not been carried into the estimation of error-bars in figure~9. If
we complete the luminosity function of SHADES galaxies towards lower
luminosities, the FIR star formation rate traced by ultraluminous
starbursts is still a factor of 1.2 to 2 lower than that of optical/UV
starbursts that have been corrected for intrinsic dust extinction.

The recent demonstration that the contribution of luminous and
ultraluminous IR galaxies dominates the SFR density at $z\le 1$ (Le
Floc'h et al. 2005) suggests that even if we correct for the incomplete
sampling of the sub-mm galaxy luminosity function, a bright SHADES
sub-mm survey could be missing the integrated contribution of dusty
starbursts to the global SFR by a large factor (as high as $\sim 7$),
and thus dusty starbursts could indeed prove to be a significant 
mode of the star formation of the Universe (Blain et al. 1999).

The contribution of the fainter ($< 3$\,mJy) 850$\mu$m sub-mm galaxy
population to the star formation history of the Universe at $z>1$
remains unconstrained at present, since detecting faint sub-mm
galaxies has been restricted to a few strongly-lensed fields
(e.g. Smail, Ivison \& Blain 1997) and extremely-deep confusion-limited pencil
observations (e.g. Hughes et al. 1998). Our estimations presented in
figure~9 should therefore be considered an educated estimate of how
ultraluminous IR-submm galaxies trace the star formation history of
the Universe.

Accurate measurements of the surface density and redshift distribution
of the entire sub-mm galaxy population (and the faintest galaxies in
particular) that contribute the complete sub-mm to FIR extragalactic
background require deeper and larger mm and sub-mm surveys than are
currently possible.
The anticipated continuum and spectroscopic
surveys with SCUBA-2
(Holland et al. 2006), the 
Large Millimetre Telescope (LMT, Serrano et al. 2006), 
and the Atacama Large Millimeter Array (ALMA, Beasley, Murowinski \& 
Tarenghi 2006), for
example, will provide suitable data.  In the meantime, however, it is
still possible that the redshift distribution of the more populous and
fainter (possibly extremely high-redshift or alternatively less
luminous) sub-mm galaxies is significantly different to those galaxies
identified in the SHADES survey.

\section{Summary of  conclusions}

\begin{itemize}

\item We have derived the photometric redshift distribution of SHADES
sources with de-boosted 850$\mu$m flux densities $> 3$~mJy towards the Lockman
Hole (LH) and the Subaru/XMM Newton Deep Field (SXDF) using rest-frame
radio to FIR photometry.  The redshift distribution of the
radio-detected sub-mm sources peaks at $z\sim 2.4$ with 50~per cent
of the population between redshifts 1.8 and 3.1.

\item The combined redshift distribution of SHADES sources with robust
radio counterparts, $\sim 60$~per cent of the population, has a
distribution which is qualitatively consistent with the distribution
of rest-frame UV-optical spectroscopic redshifts published by Chapman
et al. (2005).

\item We find a small ($\delta z \approx 0.5$), but significant,
difference between the peaks of the photometric-redshift distributions
of the LH and SXDF, which can be attributed to differences in the
sensitivities of their respective radio maps.  Intrinsic
field-to-field redshift variance is also expected, and is
characterized according to a variety of models ($\delta z \approx
0.25-0.55$).  This drives us to the conclusion that the incomplete
area ($\sim 720$ sq. arcmins) observed by SCUBA, despite being the
largest sub-mm survey to date, may still be too small to be a representative 
sample of the bright sub-mm galaxy population.  

\item The complete redshift distribution of all SHADES sources,
including those sub-mm sources without detections at radio
wavelengths (for which we adopt a variety of possibilities that
describe their unconstrained distribution of redshifts) still
maintains the peak (mode) of the bright sub-mm galaxy redshift
distribution at $\approx 2.4$. 
We have considered a variety of priors
that describe the unconstrained redshift distributions of the sub-mm
sources without radio detections. In the most extreme solutions,
distributing these sources with equal probability between their lower
redshift-limit and $z=2$ or $z=5$, the bulk of the sub-mm population
(50~per cent interquartile) lies in the range $1.6\le z\le 2.6$ or
$2.1\le z\le3.4$, respectively.

\item The combined SHADES LH and SXDF redshift probability-density 
distribution is compatible, within the 
uncertainties of our analysis, with the semi-analytical model for the
joint formation of spheroids and QSOs of Granato et al. (2004)
and Silva et al. (2005). If sources detected only at 
850$\mu$m are also introduced into the redshift probability-density, 
with other priors 
than those illustrated here, then the hydrodynamical model of Muanwong et
al. (2002) and phenomenological model of van Kampen (2004) and van Kampen et 
al. (2005) could also be in agreement with the observations.
These compatible models, which are physically quite
distinct, predict different clustering properties for the SHADES
galaxies that could allow further discrimination between them  (van Kampen et
al. 2005). A detailed study of the clustering properties of SHADES
galaxies will be the topic of a further paper.

\item The bright SHADES galaxies contribute to the SFR density of the
Universe with $\sim 0.01$ to 0.03~\Msun~yr$^{-1}$~Mpc$^{-3}$ in the
redshift interval $1 \lsim z \lsim 5$, and reach the levels of the
dust-uncorrected Lyman Break Galaxy population (Giavalisco et
al. 2004). The SFR density of dust-enshrouded starburst galaxies traced by
ultraluminous SHADES galaxies, and completing the luminosity function
to lower luminosity galaxies, is estimated to be a factor of 2
larger. This is  still a factor of 1.2 to 2 lower than the
optical/UV-selected starburst galaxy samples that include the latest
dust-correction estimates. The current SHADES survey and complementary
multiwavelength data, however, cannot
characterize the bulk of the rest-frame FIR emission arising from
these lower luminosity galaxies. A more statistically-complete
measurement of the universal history of star-formation from powerful
dusty, optically-obscured galaxies awaits the commissioning of future
large-aperture single-dish and interferometric submillimetre and
millimetre telescopes targetting suitable extragalactic 
fields that have the necessary multi-wavelength ancillary data.

\end{itemize}

\section*{Acknowledgments} 
IA and DHH acknowledge partial support from
Conacyt grants 39548-F and 39953-F and a Royal Society -- Academia
Mexicana de Ciencias 2005 exchange program. This work has been also
partially funded by PPARC. KC acknowledges additional partial support from 
NSERC, IRS from the
Royal Society, AJB from the Gordon and Betty Moore Foundation, MS from
a Gary McCue Fellowship through the Center of Cosmology at UC Irvine,
and CS from a PPARC Advanced Fellowship.

\newpage

\begin{table*}
 \begin{minipage}{170mm}
\begin{center}

\caption{Photometric redshifts for SHADES sources in the 
LH field based on the 1.4GHz/850$\mu$m spectral index.  The columns
give: (1) name of the source; (2) $z_{\rm phot}^{\rm CY}$, photometric
redshift using the prescription of Carilli \& Yun (1999, 2000); (3)
$z_{\rm phot}^{\rm A}$, photometric redshift using the template
collection of Aretxaga et al. (2003, 2005),  the 90\%
confidence interval is given in parenthesis;  (4) notes on which
radio counterpart (from Ivison et al. 2007) 
is used in the computation of photo-$z$, in case of
ambiguity (N for Northern component, S for Southern component, etc, 
or `coadded' if the flux densities from all components are summed);
 and (5) $z_{\rm spec}$, spectroscopic redshift taken from
the literature, where sources for which the redshifts are in parenthesis 
have reported
ambiguities in their radio/optical counterpart associations, or where  
the redshifts are otherwise  under scrutiny.
The references for the spec-$z$ (as a superscript of the values) and
any debate about them (after the parenthesis, where it applies) are as follows: 1.-
Chapman et al. 2005; 2.- Ivison et al. 2005; 3.- Greve et al. 2005;
4.- Chapman et al. 2003; 5.- Swinbank et al. 2005; 6.- Chapman et
al. 2002; 7.- Almaini et al. 2005; 8.- Kov\'acs et al. 2006.  }
\label{tab:CY_LH}
\begin{tabular}{lccll}
\hline
   object & $z_{\rm phot}^{\rm CY}$ &  $z_{\rm phot}^{\rm A}$ & notes & $z_{\rm spec}$ \\
\hline
SHADES J105201+572443 (Lock850.1)  & $3.3\pm_{1.2}^{1.8}$ & $2.1\pm_{0.1}^{2.0}$ (1.5--6.0) &  & (2.148$^{1,2}$)$^{2}$\\ 
%           (tentative: multicomponent$^2$)
SHADES J105257+572105 (Lock850.2)  & $5.1\pm_{2.1}^{3.5}$ & $3.1\pm_{0.1}^{2.7}$ (2.2--7.0) &  SW & \\
         & $6.0\pm_{2.5}^{4.0}$ & $3.6\pm_{0.6}^{1.7}$ (2.5--6.9) &  NW  & \\
SHADES J105257+572105 (Lock850.3)  & $7.0\pm_{3.3}^{4.4}$ & $4.1\pm_{0.6}^{1.6}$ (2.7--7.4) &  S  
& (3.036$^{1}$)$^2$\\
% taken out: no CO$^3$\\
         & $4.2\pm_{1.6}^{2.8}$ & $3.1\pm_{0.8}^{1.6}$ (2.0--6.5) &  coadded  & \\
SHADES J105204+572658 (Lock850.4)  & $3.1\pm_{1.1}^{1.6}$ & $2.1\pm_{0.4}^{1.8}$ (1.5--5.8) &  coadded  & (0.526 or 1.482)$^2$\\
SHADES J105302+571827 (Lock850.5)  & $\ge3.8$ & $\ge2.9$  &  & \\
SHADES J105204+572526 (Lock850.6)  & $7.3\pm_{3.6}^{4.5}$ & $4.1\pm_{0.6}^{1.5}$ (3.0--7.8)  & & \\
SHADES J105301+572554 (Lock850.7)  & $4.4\pm_{1.8}^{3.0}$ & $3.1\pm_{0.6}^{1.8}$ (2.0--6.7)  & & \\
SHADES J105153+571839 (Lock850.8)  & $\ge 3.0$ & $\ge2.5$  & & \\
SHADES J105216+572504 (Lock850.9)  & $3.3\pm_{1.2}^{1.8}$ & $2.1\pm_{0.1}^{2.0}$ (1.5--5.9) &  & 1.85$^2$\\
SHADES J105248+573258 (Lock850.10) & $6.4\pm_{2.1}^{0.7}$  & $3.4\pm_{0.6}^{2.3}$ (2.2--7.3)  & & \\
SHADES J105129+572405 (Lock850.11) & $\ge 2.7$ & $\ge2.3$  &  & \\
SHADES J105227+572513 (Lock850.12) & $3.6\pm_{1.5}^{2.2}$ & $2.6\pm_{0.6}^{1.7}$ (1.5--6.5) &  & (2.142$^1$)$^2$ \\
% (IS abs, not wholly convincing$^2$)\\
SHADES J105132+573134 (Lock850.13) & $\ge 2.6$ & $\ge1.5$ & & \\
%SHADES J105230+572215 (Lock850.14) & $4.4\pm_{1.7}^{2.9}$ & $3.1\pm_{0.9}^{1.6}$ (1.8--6.2) & $3.6\pm_{0.7}^{0.6}$ (2.5--5.0) & adopted $P=0.068$ & 2.611$^{4,2}$\\
SHADES J105230+572215 (Lock850.14) & $\ge 4.0$ & $\ge 2.2$ & no 1.4GHz\footnote[1]{This source has a robust 1.4GHz association in
  the dataset of Ivison et al (2002), but it is below the
robustness level adopted  for the analysis in this paper, and thus we will 
make use of the revised 1.4GHz photometry of Ivison et al. (2007) as an 
upper limit.} & 2.611$^{4,2}$\\
% taken out:, no CO$^3$\\
% without radio, A: 4.1  1.4  1.8  2.2  7.4
% without radio, HM:4.9  1.3  1.4  3.0  7.4
% without radio, CY:5.6  0.7  1.9  4.0  8.0  6.1 SHADES_LH850.14 
SHADES J105319+572110 (Lock850.15) & $3.5\pm_{1.4}^{2.0}$ & $2.6\pm_{0.8}^{1.6}$ (1.1--6.0) &  coadded  &   \\
         & $4.6\pm_{2.1}^{3.2}$ & $3.1\pm_{0.9}^{1.9}$ (1.7--6.9) & S & \\
SHADES J105151+572637 (Lock850.16) & $2.3\pm_{0.8}^{1.3}$ & $1.6\pm_{0.4}^{1.3}$ (1.0--4.7) &  & (1.147$^{1}$)$^2$\\
% (muddled ID?$^2$)\\
SHADES J105158+571800 (Lock850.17) & $2.3\pm_{0.8}^{1.2}$ & $1.6\pm_{0.4}^{1.2}$ (1.0--4.2) &  & 2.239$^{1,2,3,5}$\\
%taken out:, CO$^3$\\
SHADES J105227+572217 (Lock850.18) & $4.5\pm_{1.8}^{3.2}$ & $3.1\pm_{0.6}^{2.1}$ (1.5--6.4) &  & (1.956$^{1}$)$^4$ \\
% (~4'' offset)$^2$\\
SHADES J105235+573119 (Lock850.19) & $\ge 2.4$ & $\ge1.7$  & & \\
SHADES J105256+573038 (Lock850.21) & $\ge 2.0$ & $\ge1.5$   & & \\
SHADES J105137+573323 (Lock850.22) & $\ge 2.8$ & $\ge2.0$   & & \\
SHADES J105213+573154 (Lock850.23) & $\ge 2.4$ & $\ge1.6$   & & \\
SHADES J105200+572038 (Lock850.24) & $3.0\pm_{1.3}^{1.7}$ & $2.6\pm_{1.1}^{1.2}$ (1.1--5.8) & & \\
SHADES J105240+572312 (Lock850.26) & $4.3\pm_{2.0}^{3.0}$ & $3.1\pm_{1.1}^{1.8}$ (1.5--7.3) & & \\
SHADES J105203+571813 (Lock850.27) & $5.1\pm_{2.3}^{3.5}$ & $3.4\pm_{1.1}^{1.6}$ (2.0--6.7)  &  & \\
SHADES J105257+573107 (Lock850.28) & $\ge 2.6$ & $\ge2.0$  &  & \\
SHADES J105130+572036 (Lock850.29) & $\ge 2.8$ & $\ge 2.2$ &   & \\
SHADES J105207+571906 (Lock850.30) & $1.5\pm_{0.6}^{0.8}$ & $1.1\pm_{0.4}^{0.8}$ (0.5--3.2) & & 2.692$^1$ \\
SHADES J105216+571621 (Lock850.31) & $3.7\pm_{1.5}^{2.2}$ & $2.6\pm_{0.6}^{1.9}$ (1.5--6.6) & & \\
SHADES J105155+572311 (Lock850.33) & $2.7\pm_{1.1}^{1.3}$  & $1.9\pm_{0.6}^{1.2}$ (1.2--5.2) & &            (3.699$^{1}$, 2.686$^{4,2}$)\\
% maybe:                 2.686$^{2}$ (3.699$^{1,4}$ rejected)\\
%      
SHADES J105213+573328 (Lock850.34) & $4.9\pm_{0.9}^{0.7}$  & $3.4\pm_{1.0}^{1.6}$ (2.0--6.5)  &  & \\
SHADES J105246+572056 (Lock850.35) & $\ge 2.9$ & $\ge2.0$  &  & \\
SHADES J105209+571806 (Lock850.36) & $\ge 3.4$ & $\ge2.8$  & & \\
SHADES J105124+572334 (Lock850.37) & $4.4\pm_{1.4}^{0.9}$ & $2.9\pm_{1.1}^{1.6}$  (1.2--6.3)   & N ($P=0.013$)& \\
          & $7.1_{1.6}^{1.8}$ & $3.9\pm_{1.1}^{2.6}$ (2.0--7.8)  & adopted $P=0.078$& \\
SHADES J105307+572431 (Lock850.38) & $4.2\pm_{1.6}^{2.1}$ & $2.4\pm_{1.1}^{1.6}$ (1.2--6.2)  & & \\
SHADES J105224+571609 (Lock850.39) & $\ge 3.1$ & $\ge2.5$  &  & \\
SHADES J105202+571915 (Lock850.40) & $4.3\pm_{2.0}^{3.1}$ & $2.6\pm_{0.6}^{2.0}$ (1.5--6.4) &  & \\
SHADES J105159+572423 (Lock850.41) & $2.9\pm_{1.1}^{1.5}$ & $2.1\pm_{0.6}^{1.4}$ (1.3--5.3) &  S  & (0.689$^{1}$)$^{6,2,7,8}$ \\
% (lensed?$^6$,           2'' offset$^2$ )\\
  &  $2.4\pm_{0.4}^{0.3}$ & $1.4\pm_{0.1}^{1.4}$ (1.0--4.7) &  N+S & \\
SHADES J105257+572351 (Lock850.43) & $4.4\pm_{2.1}^{3.2}$ & $3.1\pm_{1.1}^{1.9}$ (1.5--7.3) &   adopted $P=0.060$ & \\
SHADES J105235+572514 (Lock850.47) & $\ge 2.0$ & $\ge1.5$  &  & \\
SHADES J105256+573245 (Lock850.48) & $3.1\pm_{0.6}^{1.1}$ & $2.4\pm_{0.9}^{1.6}$ (1.2--6.2) &  adopted $P=0.068$ & \\
SHADES J105245+573121 (Lock850.52) & $3.1\pm_{1.4}^{2.0}$ & $2.6\pm_{1.1}^{1.6}$ (0.5--5.7) &  & \\
SHADES J105240+571928 (Lock850.53) & $\ge 2.3$ & $\ge1.5$  &  & \\
\hline
\end{tabular}
\end{center}
\end{minipage}
\end{table*}

\setcounter{table}{0}

\begin{table*}
 \begin{minipage}{170mm}
\begin{center}
\caption{ (cont.)}
\begin{tabular}{lccll}
\hline
  object & $z_{\rm phot}^{\rm CY}$ &  $z_{\rm phot}^{\rm A}$ & 
 notes & $z_{\rm spec}$ \\
\hline
SHADES J105143+572446 (Lock850.60) & $\ge 1.4$ & $\ge0.8$  &  & \\
SHADES J105153+572505 (Lock850.63) & $3.9\pm_{1.7}^{2.6}$ & $2.6\pm_{0.6}^{2.0}$ (1.5--6.7) & & \\
SHADES J105251+573242 (Lock850.64) & $\ge 2.3$ & $\ge1.5$  &  & \\
SHADES J105138+572017 (Lock850.66) & $\ge 2.6$ & $\ge2.0$  &  & \\
SHADES J105209+572355 (Lock850.67) & $\ge 1.7$ & $\ge1.0$  &  & \\
SHADES J105148+573046 (Lock850.70) & $\ge 1.9$ & $\ge1.5$  &  & \\   
SHADES J105218+571903 (Lock850.71) & $2.1\pm_{0.8}^{1.1}$ & $1.6\pm_{0.6}^{1.1}$ (0.8--4.2) & & \\
SHADES J105141+572217 (Lock850.73) & $3.5\pm_{1.7}^{2.3}$ & $2.6\pm_{0.6}^{1.1}$ (1.0--6.2) & N & \\
         & $2.5\pm_{1.0}^{1.5}$ & $2.1\pm_{1.1}^{1.1}$ (0.5--5.0) &  coadded & \\
SHADES J105315+572645 (Lock850.75) & $\ge 1.2$ & $\ge 1.1$ &    & \\
SHADES J105148+572838 (Lock850.76) & $3.0\pm_{1.3}^{1.8}$ & $2.1\pm_{0.9}^{1.9}$ (0.8--6.0) &  & \\
SHADES J105157+572210 (Lock850.77) & $4.5\pm_{2.4}^{3.1}$ & $3.1\pm_{0.6}^{1.0}$ (1.5--7.0) &  S & \\
          & $2.4\pm_{0.5}^{0.4}$ & $1.9\pm_{0.8}^{1.1}$ (0.8--4.8) &   S+N & \\
SHADES J105145+571738 (Lock850.78) & $\ge 1.9$ & $\ge 1.3$  &  & \\
SHADES J105152+572127 (Lock850.79) & $3.6\pm_{1.6}^{2.4}$ & $2.6\pm_{0.6}^{2.0}$ (1.2--6.5) &  adopted $P=0.064$ & \\
SHADES J105231+571800 (Lock850.81) & $\ge 2.2$ & $\ge 1.9$  &  \\
SHADES J105307+572839 (Lock850.83) & $\ge 2.1$ & $\ge 0.8$ &  & \\
SHADES J105153+571733 (Lock850.87) & $2.1\pm_{0.8}^{1.1}$ & $1.6\pm_{0.6}^{1.1}$ (0.5--4.2) & & \\
SHADES J105139+571509 (Lock850.100) &$\ge 4.0$ & $\ge3.0$  &  & \\
\hline
\end{tabular}
\end{center}
\end{minipage}
\end{table*}

\setcounter{table}{1}

\begin{table*}
 \begin{minipage}{170mm}
\begin{center}

\caption{Same as table~1 for SHADES sources in the SXDF field. }
\label{tab:CY_SXDF}
\begin{tabular}{lccll}
\hline
  object & $z_{\rm phot}^{\rm CY}$ &  $z_{\rm phot}^{\rm A}$ & 
 notes & $z_{\rm spec}$ \\
\hline
SHADES J021730-045937 (SXDF850.1)  &   $4.3\pm_{1.7}^{2.9}$  & $3.1\pm_{0.6}^{2.0}$  (2.0--6.8)  & & \\ 
SHADES J021803-045527 (SXDF850.2)  &   $2.6\pm_{1.0}^{1.4}$  & $2.6\pm_{1.1}^{0.7}$  (1.2--5.1)  & & \\ 
SHADES J021742-045628 (SXDF850.3)  &   $3.3\pm_{1.3}^{1.8}$  & $2.1\pm_{0.4}^{1.8}$  (1.5--6.0)  & & \\ 
SHADES J021738-050337 (SXDF850.4)  &   $1.7\pm_{0.6}^{0.9}$ & $1.1\pm_{0.3}^{1.1}$  (0.5--3.5)  & & \\ 
SHADES J021802-050032 (SXDF850.5)  &   $1.4\pm_{0.5}^{0.6}$ & $1.1\pm_{0.5}^{0.6}$  (0.5--2.8)  & & \\ 
SHADES J021729-050326 (SXDF850.6)  &   $3.1\pm_{0.5}^{0.6}$ & $2.4\pm_{0.6}^{1.9}$  (1.3--6.0)  & NW & \\ 
           &   $3.6\pm_{1.5}^{1.1}$ & $2.9\pm_{1.1}^{1.6}$  (1.6--6.5) &  N   &\\ 
SHADES J021738-050523 (SXDF850.7)  &   $4.1\pm_{1.7}^{2.8}$ & $2.6\pm_{0.6}^{1.9}$  (2.0--7.2) &  & \\ 
SHADES J021744-045554 (SXDF850.8)  &   $3.3\pm_{1.3}^{1.9}$ & $2.6\pm_{0.7}^{1.6}$  (1.2--5.9) &  & \\ 
SHADES J021756-045806 (SXDF850.9)  &   $\ge2.1$  & $\ge1.6$  &  & \\ 
SHADES J021825-045557 (SXDF850.10) &   $2.3\pm_{0.8}^{1.3}$ & $1.6\pm_{0.4}^{1.4}$  (1.0--4.8) &  & \\ 
SHADES J021725-045937 (SXDF850.11) &   $2.8\pm_{1.2}^{1.5}$ & $2.1\pm_{0.6}^{1.6}$  (1.0--5.6) &  & \\ 
SHADES J021759-050503 (SXDF850.12) &   $3.6\pm_{1.5}^{2.2}$ & $2.6\pm_{0.9}^{1.7}$  (1.5--6.7) &  & \\ 
SHADES J021819-050244 (SXDF850.14) &   $3.4\pm_{1.6}^{2.0}$ & $2.6\pm_{1.0}^{1.6}$  (1.2--6.5) &  & \\ 
SHADES J021815-045405 (SXDF850.15) &   $\ge2.5$ & $\ge2.0$  &  & \\ 
SHADES J021813-045741 (SXDF850.16) &   $3.5\pm_{1.6}^{2.1}$ & $2.6\pm_{0.6}^{1.9}$  (1.5--6.5) &  & \\ 
SHADES J021754-045302 (SXDF850.17) &   $\ge2.7$ & $\ge2.2$  &  & \\ 
SHADES J021757-050029 (SXDF850.18) &   $3.9\pm_{1.7}^{2.5}$ & $2.6\pm_{0.6}^{2.0}$  (1.5--6.5) & & \\ 
SHADES J021828-045839 (SXDF850.19) &   $2.2\pm_{0.8}^{1.2}$ & $1.6\pm_{0.6}^{1.1}$  (0.8--4.6) &  & \\ 
SHADES J021744-050216 (SXDF850.20) &   $\ge1.7$ & $\ge1.4$  &  & \\ 
SHADES J021742-050427 (SXDF850.21) &   $0.9\pm_{0.4}^{0.6}$ & $0.6\pm_{0.2}^{0.9}$  (0.0--2.2) &  & \\ 
SHADES J021800-050741 (SXDF850.22) &   $\ge1.7$ & $\ge1.8$  &  & \\ 
SHADES J021742-050545 (SXDF850.23) &   $2.7\pm_{1.0}^{1.5}$ & $2.1\pm_{0.6}^{1.4}$  (1.0--5.0) &  & \\ 
SHADES J021734-050437 (SXDF850.24) &   $3.4\pm_{1.7}^{2.0}$ & $2.1\pm_{0.6}^{1.8}$  (1.0--6.1) &  N  & \\ 
           &   $3.7\pm_{1.8}^{2.4}$ & $2.6\pm_{0.6}^{2.0}$  (1.5--7.0) &  S  & \\ 
SHADES J021812-050555 (SXDF850.25) &   $\ge1.6$ & $\ge1.0$  &  & \\ 
SHADES J021807-050148 (SXDF850.27) &   $1.5\pm_{0.6}^{0.8}$ & $1.1\pm_{0.5}^{0.9}$  (0.5--3.3) &   & \\ 
SHADES J021807-045915 (SXDF850.28) &   $2.3\pm_{0.9}^{1.3}$  & $1.6\pm_{0.6}^{1.4}$  (0.5--4.5) &  N & \\ 
           &   $1.7\pm_{0.6}^{1.0}$ & $1.1\pm_{0.3}^{1.1}$  (0.2--3.5)  & N+S  & \\ 
SHADES J021816-045511 (SXDF850.29) &   $1.6\pm_{0.6}^{0.9}$ & $1.1\pm_{0.4}^{0.9}$  (0.6--3.5) & & \\ 
SHADES J021740-050116 (SXDF850.30) &   $4.4\pm_{2.0}^{3.1}$ & $3.1\pm_{1.1}^{1.7}$  (1.8--7.7) & & \\ 
SHADES J021736-045557 (SXDF850.31) &   $3.2\pm_{1.3}^{1.8}$ & $2.1\pm_{0.6}^{1.6}$  (1.5--6.2) & & \\ 
SHADES J021722-050038 (SXDF850.32) &   $\ge2.1$ & $\ge1.5$  &  & \\ 
SHADES J021800-045311 (SXDF850.35) &   $3.3\pm_{1.5}^{2.0}$ & $2.1\pm_{0.5}^{2.1}$  (1.0--6.1) &  & \\ 
SHADES J021832-045947 (SXDF850.36) &   $\ge1.9$ & $\ge1.8$  &  & \\ 
SHADES J021724-045839 (SXDF850.37) &   $3.2\pm_{1.5}^{2.0}$ & $2.1\pm_{0.5}^{2.1}$  (0.8--6.1) & & \\ 
SHADES J021825-045714 (SXDF850.38) &   $2.7\pm_{1.3}^{1.6}$ & $2.1\pm_{0.8}^{1.6}$  (0.5--5.4) & & \\ 
SHADES J021750-045540 (SXDF850.39) &   $\ge1.7$ & $\ge1.5$  &  & \\ 
SHADES J021729-050059 (SXDF850.40) &   $2.9\pm_{1.4}^{1.6}$   & $2.1\pm_{0.6}^{2.0}$  (1.0--5.9) & & \\ 
SHADES J021829-050540 (SXDF850.45) &   $\ge4.4$ & $\ge3.3$  &  \\ 
SHADES J021733-045857 (SXDF850.47) &   $1.5\pm_{0.6}^{0.9}$ & $1.1\pm_{0.6}^{0.8}$  (0.2--3.1) & NE & \\ 
           &   $2.6\pm_{1.1}^{1.6}$ & $2.1\pm_{1.0}^{1.4}$  (0.6--5.2) & SE  & \\ 
           &   $1.3\pm_{0.5}^{0.7}$ & $1.1\pm_{0.6}^{0.7}$  (0.2--2.9) & NE+SE & \\ 
           &   $2.2\pm_{1.0}^{1.2}$ & $1.6\pm_{0.6}^{1.3}$  (0.5--4.4) & W  & \\ 
SHADES J021724-045717 (SXDF850.48) &   $\ge2.6$ & $\ge1.8$   &  & \\ 
SHADES J021820-045648 (SXDF850.49) &   $\ge1.3$ & $\ge1.0$   &  & \\ 
SHADES J021802-045645 (SXDF850.50) &   $3.6\pm_{1.7}^{2.3}$ & $2.1\pm_{0.6}^{2.1}$  (1.2--6.9) & & \\ 
SHADES J021804-050453 (SXDF850.52) &   $2.0\pm_{0.8}^{1.1}$ & $1.6\pm_{0.6}^{1.1}$  (0.5--4.1) & E  & \\ 
           &   $1.5\pm_{0.6}^{0.9}$ & $1.1\pm_{0.6}^{0.9}$  (0.2--3.1) &  ES & \\ 
SHADES J021752-050446 (SXDF850.55) &   $3.0\pm_{1.6}^{1.7}$ & $2.1\pm_{0.6}^{2.1}$  (0.8--6.1) &  & \\ 
SHADES J021750-050631 (SXDF850.56) &   $\ge1.2$ & $\ge0.6$   &  & \\ 
SHADES J021745-045750 (SXDF850.63) &   $\ge1.6$ & $\ge1.5$   &  & \\ 
SHADES J021807-050403 (SXDF850.65) &   $\ge2.2$ & $\ge1.2$   &  & \\ 
SHADES J021751-050250 (SXDF850.69) &   $\ge1.5$ & $\ge1.0$   &  & \\ 
SHADES J021811-050247 (SXDF850.70) &   $\ge1.9$ & $\ge1.2$   &  & \\ 
SHADES J021821-045903 (SXDF850.71) &   $\ge1.7$ & $\ge1.2$   &  & \\ 
SHADES J021758-045428 (SXDF850.74) &   $2.9\pm_{1.5}^{1.7}$ & $2.1\pm_{1.1}^{1.6}$  (0.8--5.8) &  & \\ 
SHADES J021755-050621 (SXDF850.76) &   $\ge1.9$ & $\ge1.5$  &  & \\ 
SHADES J021736-050432 (SXDF850.77) &   $2.9\pm_{1.5}^{1.7}$ & $2.1\pm_{0.6}^{2.1}$  (1.0--6.1) && \\ 
SHADES J021817-050404 (SXDF850.86) &   $\ge1.5$ & $\ge1.0$  & & \\ 
SHADES J021800-050448 (SXDF850.88) &   $\ge1.0$  & $\ge 1.0$  &   & \\ 
SHADES J021734-045723 (SXDF850.91) &   $\ge1.5$ & $\ge1.2$   &  & \\ 
SHADES J021733-045813 (SXDF850.93) &   $\ge1.9$ & $\ge0.8$   &  & \\ 
\hline
\end{tabular}
\end{center}
\end{minipage}
\end{table*}

\setcounter{table}{1}

\begin{table*}
 \begin{minipage}{170mm}
\begin{center}
\caption{ (cont.)}
\begin{tabular}{lcccll}
\hline
  object & $z_{\rm phot}^{\rm CY}$ &  $z_{\rm phot}^{\rm A}$ & 
 notes & $z_{\rm spec}$ \\
\hline
SHADES J021740-045817 (SXDF850.94)  & $\ge1.5$ & $\ge1.2$   &  & \\ 
SHADES J021741-045833 (SXDF850.95)  & $\ge1.6$ & $\ge1.0$   &  & \\ 
SHADES J021800-050212 (SXDF850.96)  & $3.5\pm_{1.6}^{2.1}$ & $3.1\pm_{1.5}^{1.1}$  (1.0--6.3) &  & \\ 
SHADES J021756-045255 (SXDF850.119) & $3.4\pm_{1.6}^{2.0}$ & $2.6\pm_{1.1}^{1.7}$  (0.8--6.2) &  & \\ 
\hline
\end{tabular}
\end{center}
\end{minipage}
\end{table*}

\begin{table*}
 \begin{minipage}{170mm}
\begin{center}
\caption{Photometric redshifts for the sources with additional 
radio-mm-FIR data reported in the literature from other surveys. The
photometry has been complemented at other wavelengths, while
preserving the SHADES 450/850$\mu$m (Coppin et al. 2006) and 1.4GHz
(Ivison et al. 2007) determined flux densities.  The first column gives the
source-name; the second column gives the most probable mode and error bars
based on the 68\% confidence interval of the mode calculation (in
parenthesis the 90\% confidence interval); the third and fourth columns
respectively give the bands in which the source is detected
at a $\ge 3\sigma$ level, and at which 
upper-limits are used for the 
computation of the photometric-redshifts;
the fifth column provides the published references (and alternative 
sources-names in overlapping surveys) to the additional photometry, 
using the following syntax:
LE850.x from Scott et al. 2002, 
1100.x from Laurent et al. 2005, 2006 and 1200.x from Greve et al. 2004 and Ivison 
et al. 2005).}
\label{3filters}
\begin{tabular}{lcccl}
\hline
  object & $z_{\rm phot}$  & $\geq
  3\sigma$ detections & $< 3\sigma$ / upper limits & notes \\
\hline
Lock850.1 & 2.4$\pm_{0.2}^{1.1}$ (2.2--3.8) & 350,850$\mu$m,1.1,1.2mm,1.4GHz & 175,450$\mu$m,3.3mm,5GHz & LE850.1, 1200.5, 1100.14 
\\
Lock850.2 & 2.9$\pm_{0.1}^{0.3}$ (2.5--3.8) & 350,850$\mu$m,
1.1,1.2mm, 1.4GHz & 450$\mu$m  & LE1100.1, 1200.4, SW-1.4GHz  \\
        & 2.9$\pm_{0.1}^{0.7}$ (2.8--3.9)    &    &   & \ \ \ \ \ \ \ \ \ \ \ \ \ \ \ \ \ \ \ \ \ \ \ \  NW-1.4GHz   \\
Lock850.3 & 2.9$\pm_{0.3}^{0.9}$  (2.5--4.2)    & 350,850$\mu$m,
1.1,1.2mm, 1.4GHz & 450$\mu$m, 5GHz  & LE850.2, 1100.8, 1200.1, S-1.4GHz \\
        & 2.6$\pm_{0.1}^{0.3}$  (2.5--3.8)    &    &   & \ \ \ \ \ \ \ \ \ \ \ \ \ \ \ \ \ \ \ \ \ \ \ \ coadded-1.4GHz \\
Lock850.4 & 1.6$\pm_{0.1}^{0.3}$  (1.5--4.8)    & 850$\mu$m,
1.2mm, 1.4GHz & 450$\mu$m, 5GHz  & LE850.14, 1200.3 \\
Lock850.12 & 2.6$\pm_{0.1}^{0.2}$  (2.2--3.0)   & 350,850$\mu$m,
1.1,1.2mm, 1.4GHz & 450$\mu$m, 5GHz & LE850.16, 1100.16, 1200.6  \\
Lock850.14 & 2.6$\pm_{0.1}^{0.8}$  (2.2--3.7)   & 350,850$\mu$m,
1.1,1.2mm & 450$\mu$m, 1.4,5GHz  & LE850.6, 1100.5, 1200.10\footnote[1]{This source has a robust 1.4GHz association in
  the dataset of Ivison et al (2002), but it is below the
robustness level adopted  for the analysis in this paper, and thus we will 
make use of the 1.4GHz photometry of Ivison et al. (2007) as an 
upper limit.} \\
Lock850.16 & 1.9$\pm_{0.1}^{0.4}$  (1.5--3.2)   & 850$\mu$m,
1.2mm,1.4GHz & 450$\mu$m, 5GHz & LE850.7 \\
Lock850.17 & 2.5$\pm_{0.5}^{0.6}$  (2.0--5.9)   & 850$\mu$m,
1.2mm, 5,1.4GHz & 450$\mu$m  & LE850.3, 1200.11 \\
Lock850.18 & 3.1$\pm_{0.1}^{2.9}$  (2.3--6.0)    & 850$\mu$m,
1.2mm, 1.4GHz & 450$\mu$m  & LE1200.9\\
Lock850.27 & 4.6$\pm_{0.4}^{1.4}$  (4.0--6.0)    & 850$\mu$m,
1.1,1.2mm, 1.4GHz & 450$\mu$m, 5GHz  & LE1100.4, 1200.7  \\
Lock850.33 & $3.6\pm_{0.9}^{0.7}$ (2.4--4.8)   & 850$\mu$m,
1.2mm, 1.4GHz & 450$\mu$m  & LE850.18, 1200.12 \\
Lock850.41 & 3.4$\pm_{0.2}^{0.7}$  (3.2--4.4) & 350,850$\mu$m,
1.1,1.2mm, 1.4GHz & 450$\mu$m  & LE850.8, 1100.17, 1200.14 \\
Lock850.76 & 4.6$\pm_{1.1}^{1.4}$  (3.0--6.0) & 850$\mu$m,1.1mm,1.4GHz & 
450$\mu$m  & LE1100.15 \\
\hline
\end{tabular}
\end{center}
\end{minipage}
\end{table*}

\begin{table*}
 \begin{minipage}{170mm}
\begin{center}

\caption{Photometric redshifts for SHADES sources in the LH field
  based on the 850$\mu$m and 1.4GHz data and 450$\mu$m upper limits
  determined by SHADES.  The columns give: (1) name of the source; (2)
  $z_{\rm phot}^{\rm SA}$ survival analysis solution; (3) $z_{\rm
    phot}^{\rm MC}$ Monte Carlo solution; and (4) notes on
  associations.}
\label{tab:481_LH}
\begin{tabular}{lccl}
\hline
  object & $z_{\rm phot}^{\rm SA}$ &  $z_{\rm phot}^{\rm MC}$ & notes \\
\hline
Lock850.1  & $2.1 \pm_{0.1}^{2.0}$ (1.5--5.9) & $2.4\pm_{0.2}^{0.1}$ (2.0--2.5) & \\
Lock850.2  & $3.1 \pm_{0.1}^{2.8}$ (2.4--6.8) & $2.9\pm_{0.1}^{0.6}$ (2.6--3.8) & SW  \\
           & $3.6\pm_{0.6}^{2.0}$  (2.5--6.8) & $3.6\pm_{0.4}^{0.4}$ (3.0--4.2) & NW  \\
Lock850.3  & $4.1\pm_{0.6}^{2.3}$ (2.8--7.7)  & $3.9\pm_{0.4}^{0.6}$ (3.2--4.8) & S \\
           & $3.1\pm_{0.8}^{1.6}$ (2.0--6.1)  & $2.6\pm_{0.3}^{0.4}$ (2.2--3.2) & coadded \\
Lock850.4  & $2.1\pm_{0.4}^{1.8}$ (1.5--5.8)  & $2.1\pm_{0.1}^{0.3}$ (1.8--3.1) & coadded  \\
Lock850.5  & $\ge 2.9$ & $\ge 3.0$  & \\
Lock850.6  & $4.1\pm_{0.6}^{2.5}$ (3.0--8.0) & $3.6\pm_{0.1}^{1.0}$ (2.8--4.8) & \\
Lock850.7  & $3.1\pm_{0.6}^{1.8}$ (2.0--6.3) & $2.9\pm_{0.3}^{0.4}$ (2.2--3.4) & \\
Lock850.8  & $\ge 2.5$ &  $\ge 2.5$  & \\
Lock850.9  & $2.1\pm_{0.1}^{2.0}$ (1.5--5.8) & $2.4\pm_{0.4}^{0.3}$ (2.0--3.2) & \\
Lock850.10 & $3.4\pm_{0.6}^{2.3}$ (2.2--7.0)  & $3.1\pm_{0.3}^{0.9}$ (2.8--4.7) & \\
Lock850.11 & $\ge 2.2$ &  $\ge 2.5$  & \\
Lock850.12 & $2.6\pm_{0.6}^{1.6}$ (1.5--6.2) & $2.6\pm_{0.2}^{0.4}$ (2.2--3.8) & \\
Lock850.13 & $\ge1.5$ &  $\ge 1.5$  & \\
Lock850.14 & $\ge 2.2$ & $\ge 2.2$ & 
 no 1.4GHz\footnote[1]{This source has a robust 1.4GHz association in
  the dataset of Ivison et al (2002), but it is below the
robustness level adopted  for the analysis in this paper, and thus we will 
make use of the 1.4GHz photometry of Ivison et al. (2007) as an 
upper limit.} \\
% taken out:, no CO$^3$\\
% without radio, A: 4.1  1.4  1.8  2.2  7.4
% without radio, HM:4.9  1.3  1.4  3.0  7.4
% without radio, CY:5.6  0.7  1.9  4.0  8.0  6.1 SHADES_LH850.14 
Lock850.15 & $2.6\pm_{0.8}^{1.6}$ (1.2--5.9) & $2.4\pm_{0.4}^{0.4}$ (2.0--3.2) & coadded  \\
           & $3.1\pm_{0.9}^{2.0}$ (1.8--6.8) & $2.9\pm_{0.5}^{0.4}$ (2.2--3.8) & S \\
Lock850.16 & $1.6\pm_{0.4}^{1.3}$ (1.0--4.7) & $3.1\pm_{0.7}^{0.1}$ (2.0--3.4) & \\
Lock850.17 & $1.6\pm_{0.4}^{1.2}$ (1.0--4.2) & $2.9\pm_{0.4}^{0.3}$ (2.3--3.2)  & \\
Lock850.18 & $3.1\pm_{0.6}^{2.1}$ (1.5--6.1) & $2.9\pm_{0.5}^{0.4}$ (2.2--3.7) & \\
Lock850.19 & $\ge1.8$ &  $\ge 1.5$  & \\
Lock850.21 & $\ge1.5$ &  $\ge 1.0$ & \\
Lock850.22 & $\ge2.0$ &  $\ge 2.0$ & \\
Lock850.23 & $\ge1.6$ &  $\ge 1.5$ & \\
Lock850.24 & $2.6\pm_{1.1}^{1.2}$ (1.2--5.8) & $2.4\pm_{0.3}^{0.4}$ (2.0--3.1) & \\
Lock850.26 & $3.1\pm_{1.1}^{1.9}$ (1.5--7.2) & $3.6\pm_{0.8}^{0.1}$ (2.2--3.9) & \\
Lock850.27 & $3.9\pm_{1.1}^{1.9}$ (1.8--6.5)  &  $3.9\pm_{1.1}^{1.9}$ (1.8--6.5) & \\
Lock850.28 & $\ge2.0$  &  $\ge 2.0$ & \\
Lock850.29 & $\ge2.2$ & $\ge 2.1$ &  \\
%Lock850.29 & $3.1\pm_{0.6}^{2.3}$ (2.0--7.0) & $3.6\pm_{0.8}^{0.1}$ (2.5--4.0) & adopted $P=0.066$ & \\
Lock850.30 & $1.1\pm_{0.4}^{0.8}$ (0.5--3.2) & $2.1\pm_{0.4}^{0.1}$ (1.8--2.5) & \\
Lock850.31 & $2.6\pm_{0.6}^{1.9}$ (1.5--6.4) & $2.6\pm_{0.6}^{0.1}$ (2.0--3.1) & \\
Lock850.33 & $2.1\pm_{0.6}^{1.4}$ (1.2--5.4) & $2.1\pm_{0.4}^{0.7}$ (1.8--3.2) & \\
Lock850.34 & $3.4\pm_{1.0}^{1.6}$ (2.0--6.5)  &  $3.1\pm_{0.2}^{0.6}$ (2.6--3.8)  & \\                 
%Lock850.34 & $\ge 3.2$  &  $\ge 3.0$  & & \\                 
Lock850.35 & $\ge2.0$  &  $\ge 2.0$ & \\ 
Lock850.36 & $\ge2.5$  &  $\ge 2.5$  & \\
Lock850.37 & $2.9\pm_{1.1}^{1.6}$  &  $4.5\pm_{0.3}^{1.0}$ (4.3--5.8) & N ($P=0.013$)\\
  & $3.9\pm_{1.1}^{2.6}$  &  $4.6\pm_{0.1}^{0.6}$ (4.5--5.8) & S (adopted $P=0.078$) \\
%Lock850.37 & $\ge2.3$  &  $\ge 2.5$ & & \\
Lock850.38 & $2.6\pm_{1.1}^{1.8}$ (0.8--6.2) & $2.1\pm_{0a.1}^{0.1}$ (2.0--2.3) & \\
Lock850.39 & $\ge2.2$  &  $\ge 2.0$ & \\
Lock850.40 & $2.6\pm_{0.6}^{2.2}$ (1.2--6.3) & $2.6\pm_{0.2}^{0.6}$ (2.0--3.2) & \\
 Lock850.41 & $2.1\pm_{0.6}^{1.4}$ (1.2--5.3) & $3.6\pm_{0.4}^{0.4}$ (2.5--4.0)  & S  \\
    & $1.4\pm_{0.1}^{1.4}$ (1.0--4.7) & $2.9\pm_{0.1}^{0.5}$ (2.3--3.5) & N+S  \\
Lock850.43 & $3.6\pm_{1.6}^{1.2}$ (1.5--6.6) & $2.4\pm_{0.1}^{0.8}$ (2.2--3.8) &  adopted $P=0.060$ \\
Lock850.47 & $\ge1.2$  &  $\ge 1.5$ & \\
Lock850.48 & $2.1\pm_{1.1}^{1.5}$ (0.5--5.7) & $2.4\pm_{0.1}^{0.5}$ (2.1--3.0) & adopted $P=0.068$ \\
Lock850.52 & $2.1\pm_{0.6}^{1.9}$ (0.5--5.5) & $2.1\pm_{0.1}^{0.1}$ (1.9--2.2) & \\
Lock850.53 & $\ge1.5$  &  $\ge 1.5$ & \\
Lock850.60 & $\ge1.2$  &  $\ge 1.5$ & \\
Lock850.63 & $2.6\pm_{0.6}^{2.1}$ (1.5--6.4) & $2.6\pm_{0.4}^{0.4}$ (2.1--3.2) & \\
Lock850.64 & $\ge 1.6$  &  $\ge 1.5$  & \\
Lock850.66 & $\ge 1.5$  &  $\ge 1.5$ & \\
Lock850.67 & $\ge1.0$   &  $\ge 1.0$ & \\
Lock850.70 & $\ge1.0$   &  $\ge 1.0$ & \\   
Lock850.71 & $1.6\pm_{0.6}^{1.2}$ (0.5--4.3) & $2.9\pm_{0.7}^{0.1}$ (2.0--3.2) & \\
\hline
\end{tabular}
\end{center}
\end{minipage}
\end{table*}

\setcounter{table}{3}

\begin{table*}
 \begin{minipage}{170mm}
\begin{center}
\caption{ (cont.)}
\begin{tabular}{lccl}
\hline
  object &  $z_{\rm phot}^{\rm SA}$ & $z_{\rm phot}^{\rm MC}$ &
  notes \\
\hline
Lock850.73 & $2.1\pm_{0.2}^{2.6}$ (1.0--6.2) & $2.4\pm_{0.2}^{0.1}$ (2.0--2.5) & N \\
           & $2.1\pm_{0.9}^{1.4}$ (0.7--5.2) & $2.1\pm_{0.1}^{0.1}$ (1.9--2.5) & coadded  \\
Lock850.75 & $\ge 1.1$ & $\ge 1.2$ &  \\
Lock850.76 & $2.1\pm_{0.6}^{1.8}$ (0.8--5.5) & $2.1\pm_{0.1}^{0.2}$ (2.0--2.5) & \\
Lock850.77 & $2.9\pm_{1.1}^{1.7}$ (2.8--6.9) & $2.6\pm_{0.1}^{0.8}$ (2.2--3.8) & S \\
           & $1.5\pm_{0.4}^{1.6}$ (0.8--4.8) & $2.8\pm_{0.4}^{0.4}$ (2.0--3.2) & N+S \\
Lock850.78 & $\ge 1.1$ &  $\ge 1.5$  & \\
Lock850.79 & $2.6\pm_{0.6}^{2.2}$ (1.2--6.3) & $2.4\pm_{0.1}^{0.6}$ (2.0--3.2) &  adopted $P=0.064$ \\
Lock850.81 & $\ge 1.9$ & $\ge 2.0$ &  \\
Lock850.83 & $\ge0.8$ &  $\ge 1.5$  & \\
Lock850.87 & $1.6\pm_{0.6}^{1.0}$ (0.8--4.0) & $2.4\pm_{0.1}^{0.6}$ (1.9--3.0) & \\
Lock850.100 &$\ge2.1$  &  $\ge 2.0$ & \\
\hline
\end{tabular}
\end{center}
\end{minipage}
\end{table*}

\setcounter{table}{4}

\begin{table*}
 \begin{minipage}{170mm}
\begin{center}

\caption{Photometric redshifts for SHADES sources in the SXDF field
  based on the 850$\mu$m and 1.4GHz data and 450$\mu$m upper limits,
 and when significant, complemented with Spitzer photometry at 70 and 
160$\mu$m. Columns are as in table~4.}
\label{tab:481_SXDF}
\begin{tabular}{lccl}
\hline
  object & $z_{\rm phot}^{\rm SA}$ & $z_{\rm phot}^{\rm MC}$ &  notes \\
\hline
SXDF850.1  &   $2.9 \pm_{0.6}^{2.0}$ (1.8--6.3) & $2.6\pm_{0.3}^{0.4}$ (2.2--3.4) &  \\ 
SXDF850.2  &   $2.4 \pm_{1.1}^{0.8}$ (1.0--4.8) & $1.9\pm_{0.1}^{0.4}$ (1.8--2.9) &  \\ 
SXDF850.3  &   $2.4 \pm_{0.6}^{1.5}$ (1.5--5.9) & $2.1\pm_{0.1}^{0.3}$ (2.0--2.6) &  \\ 
SXDF850.4  &   $1.1 \pm_{0.4}^{1.0}$ (0.5--3.4) & $2.1\pm_{0.2}^{0.4}$ (1.8--2.7) &  \\ 
SXDF850.5  &   $1.4 \pm_{0.5}^{0.6}$ (0.5--2.8) & $1.4\pm_{0.4}^{0.4}$ (0.6--2.0) & 70,160$\mu$m included \\  
%SXDF850.5  &   $1.1 \pm_{0.4}^{0.6}$ (0.5--2.8) & $1.6\pm_{0.1}^{0.2}$ (1.4--2.0) & 70,160$\mu$m included ... \\  
SXDF850.6  &   $2.4\pm_{0.6}^{1.9}$ (1.3--6.0) & $2.4\pm_{0.1}^{1.2}$ (2.1--3.8) &  NW \\ 
           &   $2.9\pm_{0.9}^{1.6}$ (1.5--6.0) & $2.6\pm_{0.4}^{0.4}$ (2.2--3.7) &  N  \\ 
SXDF850.7  &   $2.4 \pm_{0.5}^{2.1}$ (1.8--6.6) & $2.4\pm_{0.1}^{0.5}$ (2.2--3.4) &  \\ 
SXDF850.8  &   $2.4 \pm_{0.6}^{1.6}$ (1.2--5.8) & $2.6\pm_{0.1}^{1.3}$ (2.3--4.0) &  \\ 
SXDF850.9  &   $\ge1.8$ &  $\ge 1.5$ & \\ 
SXDF850.10 &   $1.4 \pm_{0.1}^{1.5}$ (0.8--4.6) & $2.6\pm_{0.6}^{0.3}$ (1.9--3.2) & \\ 
SXDF850.11 &   $1.9 \pm_{0.6}^{1.5}$ (1.0--5.6) & $2.4\pm_{0.4}^{0.4}$ (2.0--3.4) & \\ 
SXDF850.12 &   $1.9 \pm_{0.1}^{2.4}$ (1.2--6.3) & $2.4\pm_{0.4}^{0.3}$ (2.0--3.0) & \\ 
SXDF850.14 &   $2.4 \pm_{0.9}^{1.6}$ (1.2--6.3) & $2.4\pm_{0.3}^{0.4}$ (2.0--3.1) & \\ 
SXDF850.15 &   $\ge1.8$  &  $\ge 2.0$ & \\ 
SXDF850.16 &   $2.4 \pm_{0.6}^{1.9}$ (1.2--6.2) & $2.4\pm_{0.2}^{0.6}$ (2.0--3.2) \\ 
SXDF850.17 &   $\ge2.0$  &  $\ge 2.0$  & \\ 
SXDF850.18 &   $2.4 \pm_{0.6}^{2.0}$ (1.5--6.1) & $2.9\pm_{0.6}^{0.2}$ (2.2--3.7) & \\ 
SXDF850.19 &   $1.6 \pm_{0.6}^{1.1}$ (0.8--4.6) & $2.4\pm_{0.4}^{0.4}$ (2.0--3.2) & \\ 
SXDF850.20 &   $\ge1.4$  &  $\ge 1.5$ & \\ 
SXDF850.21 &   $0.6 \pm_{0.2}^{0.9}$ (0.0--2.2) & $0.5\pm_{0.2}^{0.4}$ (0.0--1.2) & 70,160$\mu$m included\\ 
SXDF850.22 &   $\ge1.8$   &  $\ge 2.0$ & \\ 
SXDF850.23 &   $1.9 \pm_{0.6}^{1.4}$ (1.0--5.0) & $2.4\pm_{0.2}^{0.9}$ (2.0--3.5) & \\ 
SXDF850.24 &   $2.4 \pm_{0.9}^{1.6}$ (1.1--6.0) & $2.4\pm_{0.1}^{0.6}$ (2.2--3.7) & \\ 
           &   $2.9 \pm_{1.1}^{1.5}$ (1.5--6.8) & $2.9\pm_{0.6}^{0.1}$ (2.2--3.7) & S \\ 
SXDF850.25 &   $\ge1.0$  &  $\ge 1.0$ & \\ 
SXDF850.27 &   $1.1 \pm_{0.5}^{0.9}$ (0.5--3.3) & $3.6\pm_{0.8}^{0.1}$ (2.4--3.8) & \\ 
SXDF850.28 &   $1.6 \pm_{0.4}^{1.1}$ (0.5--4.4) & $1.4\pm_{0.5}^{0.5}$ (0.6--2.2)  & N, 70,160$\mu$m included \\ 
           &   $1.2 \pm_{0.4}^{1.1}$ (0.2--3.2) & $1.1\pm_{0.1}^{0.4}$ (0.5--1.5) & N+S \\ 
SXDF850.29 &   $1.1 \pm_{0.4}^{0.9}$ (0.6--3.5) & $2.1\pm_{0.3}^{0.1}$ (1.8--2.4) & \\ 
SXDF850.30 &   $2.9 \pm_{1.1}^{1.8}$ (1.2--6.8) & $2.9\pm_{0.6}^{0.2}$ (2.2--3.8) & \\ 
SXDF850.31 &   $2.1 \pm_{0.6}^{1.6}$ (1.5--6.1) & $2.6\pm_{0.4}^{0.5}$ (2.2--3.7) & \\ 
SXDF850.32 &   $\ge1.5$  &  $\ge 1.5$ & \\ 
SXDF850.35 &   $2.4 \pm_{0.9}^{1.6}$ (1.2--6.2) & $2.4\pm_{0.1}^{0.6}$ (2.1--3.2) & \\ 
SXDF850.36 &   $\ge1.8$  &  $\ge 2.0$ & \\ 
SXDF850.37 &   $2.1 \pm_{0.5}^{2.1}$ (0.8--5.9) & $2.1\pm_{0.1}^{0.1}$ (2.0--2.4) & \\ 
SXDF850.38 &   $1.9 \pm_{0.6}^{1.8}$ (0.3--5.0) & $1.9\pm_{0.1}^{0.2}$ (1.7--2.2) & \\ 
SXDF850.39 &   $\ge1.2$  &  $\ge 1.5$ & \\ 
SXDF850.40 &   $2.1 \pm_{0.6}^{1.9}$ (1.0--5.8) & $2.4\pm_{0.4}^{0.4}$ (2.0--3.2) & \\ 
SXDF850.45 &   $\ge2.8$   &  $\ge 3.0$ & \\ 
SXDF850.47 &   $1.1 \pm_{0.6}^{0.8}$ (0.2--3.1) & $1.9\pm_{0.1}^{0.3}$ (1.6--2.2) & NE \\ 
           &   $2.1 \pm_{1.0}^{1.4}$ (0.6--5.2) & $2.1\pm_{0.1}^{0.1}$ (1.9--2.2) & SE \\ 
           &   $1.1 \pm_{0.6}^{0.7}$ (0.2--2.9) & $1.6\pm_{0.1}^{0.3}$ (1.5--2.1) & NE+SE\\ 
           &   $1.6 \pm_{0.6}^{1.3}$ (0.5--4.4) & $2.1\pm_{0.2}^{0.1}$ (1.8--2.4) & W \\ 
SXDF850.48 &   $\ge1.8$   &  $\ge 2.0$ & \\ 
SXDF850.49 &   $\ge0.5$   &  $\ge 1.0$ & \\  
SXDF850.50 &   $2.4 \pm_{0.9}^{2.0}$ (1.2--6.6) & $2.9\pm_{0.6}^{0.1}$ (2.2--3.6)& \\ 
SXDF850.52 &   $1.6 \pm_{0.6}^{1.1}$ (0.5--4.1) & $2.1\pm_{0.2}^{0.1}$ (1.6--2.2)& E \\ 
           &   $1.1 \pm_{0.6}^{0.9}$ (0.2--3.1) & $2.1\pm_{0.4}^{0.1}$ (1.5--2.2)& ES \\ 
SXDF850.55 &   $2.4 \pm_{1.1}^{1.6}$ (0.5--5.6) & $2.1\pm_{0.2}^{0.1}$ (1.8--2.2)& \\ 
SXDF850.56 &   $\ge0.6$ &  $\ge 1.0$ & \\ 
SXDF850.63 &   $\ge1.0$ &  $\ge 1.0$ & \\ 
SXDF850.65 &   $\ge1.2$ &  $\ge 1.5$ & \\ 
SXDF850.69 &   $\ge0.8$ &  $\ge 1.0$ & \\ 
SXDF850.70 &   $\ge1.0$ &  $\ge 1.0$ & \\ 
SXDF850.71 &   $\ge0.8$ &  $\ge 1.0$ & \\ 
SXDF850.74 &   $2.1 \pm_{1.1}^{1.6}$ (0.9--5.8) & $2.1\pm_{0.2}^{0.1}$ (1.8--2.2) & \\ 
SXDF850.76 &   $\ge1.2$   &  $\ge 1.5$ & \\ 
SXDF850.77 &   $2.1 \pm_{0.6}^{1.0}$ (1.0--6.0) & $2.1\pm_{0.1}^{0.1}$ (1.8--2.2) & $70,160\mu$m included\\ 
SXDF850.86 &   $\ge1.0$  &  $\ge 1.0$ & \\ 
\hline
\end{tabular}
\end{center}
\end{minipage}
\end{table*}

\setcounter{table}{4}

\begin{table*}
 \begin{minipage}{170mm}
\begin{center}
\caption{ (cont.)}
\begin{tabular}{lccl}
\hline
  object &   $z_{\rm phot}^{\rm SA}$ & $z_{\rm phot}^{\rm MC}$ &
 notes \\
\hline
SXDF850.88 &   $\ge 1.0$  & $\ge 1.0$  &  \\ 
SXDF850.91 &   $\ge1.0$   &  $\ge 1.5$ & \\ 
SXDF850.93 &   $\ge0.8$   &   $\ge 0.5$ & \\ 
SXDF850.94  & $\ge1.2$   & $\ge1.5$ & \\ 
SXDF850.95  & $\ge1.0$   & $\ge1.0$ & \\ 
SXDF850.96  & $2.4 \pm_{0.9}^{1.8}$ (1.0--6.1) & $2.4\pm_{0.1}^{0.6}$ (2.2--3.5) & \\ 
SXDF850.119 & $2.2\pm_{1.4}^{1.8}$  (0.0--4.5) & $1.9\pm_{0.5}^{0.1}$ (1.2--2.2) & 70,160$\mu$m included \\ 
\hline
\end{tabular}
\end{center}
\end{minipage}
\end{table*}

%%%  Already calculated 
%SXDF850.5  &    & 160,850$\mu$m,1.4GHz & 70,450$\mu$m & & \\
%SXDF850.21 &    & 70,160,850$\mu$m,1.4GHz & 450$\mu$m & & \\
%SXDF850.28 &   & 70,850$\mu$m,1.4GHz & 160,450$\mu$m & & \\

\end{document}